\documentclass[10pt,a4paper,nofootinbib,superscriptaddress]{revtex4-2}

\pdfoutput=1

\usepackage[utf8]{inputenc}
\usepackage{graphicx,bm,bbm}
\usepackage{amssymb,graphicx,epstopdf}
\usepackage{slashed,subfigure}
\usepackage{caption}
\usepackage{array}
\usepackage{xcolor}
\usepackage{amsmath,mathtools}
\usepackage{epstopdf,dcolumn}
\usepackage{soul}
\usepackage{mathtools}
\allowdisplaybreaks
\usepackage[normalem]{ulem}
\usepackage{cancel}
\usepackage{xcolor}
\usepackage[colorlinks=true,linktocpage=true,linkcolor=blue,citecolor=blue]{hyperref}
\usepackage{xcolor}
\usepackage{braket}
\allowdisplaybreaks
\usepackage{enumitem}
\usepackage{mathrsfs}
\usepackage{caption}

\def\be {\begin{equation}}
\def\ee {\end{equation}}
\def\bea {\begin{eqnarray}}
\def\eea {\end{eqnarray}}
\def\bc {\begin{center}}
	\def\ec {\end{center}}

\def\({\left(}
\def\){\right)}
\def\[{\left[}
\def\]{\right]}

\DeclareGraphicsExtensions{.jpg,.pdf,.eps}

\begin{document}

\title{Revisiting shear stress tensor evolution: Non-resistive magnetohydrodynamics with momentum-dependent relaxation time}

\author{Sunny Kumar Singh}
	\email{sunny.singh@iitgn.ac.in}
	\affiliation{Indian Institute of Technology Gandhinagar,\\ Gandhinagar-382355, Gujarat, India}

	\author{Manu Kurian}
	\email{mkurian@bnl.gov}
	\affiliation{RIKEN BNL Research Center, Brookhaven National Laboratory,\\ Upton, New York 11973, USA}

 	\author{Vinod Chandra}
	\email{vchandra@iitgn.ac.in}
	\affiliation{Indian Institute of Technology Gandhinagar,\\ Gandhinagar-382355, Gujarat, India}





\begin{abstract}

This study aims to develop second-order relativistic viscous magnetohydrodynamics (MHD) derived from kinetic theory within an extended relaxation time approximation (momentum/energy dependent) for the collision kernel. The investigation involves a detailed examination of shear stress tensor evolution equations and associated transport coefficients. The Boltzmann equation is solved using a Chapman-Enskog-like gradient expansion for a charge-conserved conformal system, incorporating a momentum-dependent relaxation time. The derived relativistic non-resistive, viscous second-order MHD equations for the shear stress tensor reveal significant modifications in the coupling with dissipative charge current and magnetic field due to the momentum dependence of the relaxation time. By utilizing a power law parametrization to quantify the momentum dependence of the relaxation time, the anisotropic magnetic field-dependent shear coefficients  in the Navier-Stokes limit have been investigated. The resulting viscous coefficients are seen to be sensitive to the momentum dependence of the relaxation time.


\end{abstract}
	
	\maketitle 

\section{Introduction}
High-energetic heavy-ion collision experiments at the Relativistic Heavy Ion Collider (RHIC) and Large Hadron Collider (LHC) have generated the hot and dense state of strongly interacting nuclear matter known as the Quark Gluon Plasma (QGP). Relativistic hydrodynamics has proven to be an effective framework for describing the space-time evolution of the QGP~\cite{Gale:2013da, Heinz:2013th}. Phenomenological analyses of bulk observables such as collective flow, hadron spectra, {\it etc}.,  highlight the importance of incorporating various dissipative processes in the hydrodynamical  evolution of the QGP~\cite{Teaney:2003kp,Romatschke:2007mq, Ryu:2015vwa}. 
Strong magnetic fields are expected to be generated at the initial phase of heavy-ion collision~\cite{Bzdak:2011yy, Deng:2012pc, Tuchin:2013ie, Huang:2015oca}. These fields may experience fast decay as the system evolves. However, some studies have shown that the decaying fields can induce electric currents within the expanding system and can contribute to the prolonged existence of magnetic fields throughout the QGP evolution. There are still uncertainties regarding the magnetic field evolution and its lifetime in the QGP medium. Recent measurements at the RHIC and LHC~\cite{Acharya:2019ijj, Adam:2019wnk}, and the related studies~\cite{Das:2016cwd, Jiang:2022uoe} on the enhanced directed flow ($v_1$) of $D$ mesons and $v_1$ splitting of $D^0$ and $\bar{D}^0$ mesons indicated the presence of a strong magnetic field in the early phase of the collision. Significant research efforts have been dedicated to explore the impact of magnetic fields on the phase diagram of Quantum Chromodynamics(QCD)~\cite{Andersen:2014xxa,Delia:2021yvk,Ayala:2021nhx}. The magnetic field appears to hold a pivotal significance, impacting not only the thermodynamic and transport properties of the QCD medium but also influencing heavy quark dynamics, dilepton spectra, and jet physics~\cite{Karmakar:2020mnj,Wang:2022jxx,Kurian:2018qwb,Ghosh:2022vjp,Dey:2020awu,Sebastian:2023tlw,Satapathy:2021cjp,Li:2017tgi,Khan:2023kkg,Gowthama:2020ghl,Rath:2020wgj,Ghosh:2023ghi,Jamal:2023ncn,Das:2022lqh}. The presence of a magnetic field has a substantial impact on the equations of motion of charged particles due to the spatial anisotropy introduced by the field. This gives rise to additional transport coefficients that depend on the orientation and strength of the magnetic field in the medium~\cite{Dash:2020vxk,Ghosh:2022xtv,Gangopadhyaya:2022olm,Rath:2023abm,Singh:2020faa,Huang:2011dc,Kurian:2021zyb,Panday:2021itx,Jaiswal:2020hvk,Goswami:2023eol,K:2021sct,Kurian:2020qjr}.  The magnetic field together with chiral anomaly of the QCD medium can lead to many interesting phenomena such as the Chiral Magnetic Effect~\cite{Fukushima:2008xe,Kharzeev:2015znc,Hattori:2016emy}. Other fascinating  phenomena associated with the magnetic field are the magnetic catalysis and inverse magnetic catalysis in the QCD medium~\cite{Mueller:2015fka, Shovkovy:2012zn}. One of the advancements in relativistic hydrodynamics is its coupling with the electromagnetic fields created during heavy-ion collisions, referred to as a magnetohydrodynamical framework~\cite{Denicol:2018rbw, Denicol:2019iyh,Huang:2009ue, Panda:2020zhr,Panda:2021pvq, Gursoy:2014aka, Inghirami:2016iru,Roy:2017yvg,Hattori:2022hyo,Hattori:2017usa,Das:2017qfi,Kushwah:2024zgd,Fang:2024skm}.

The transport coefficients of the QCD medium serve as input parameters in the magnetohydrodynamics simulation. These coefficients play a crucial role in capturing the dynamic behavior of the magnetized QCD medium within the framework of magnetohydrodynamics. The understanding of dissipative processes and their associated transport parameters relies on the non-equilibrium physics of the QGP which requires the knowledge of the microscopic description of the medium. The relativistic Boltzmann equation stands as a transport equation governing the space-time evolution of the phase-space distribution function. However, directly solving the Boltzmann equation poses a challenge due to the intricate nature of the  collision term present there. Several simplified approaches for the collision term have been proposed over the course of several decades~\cite{Bhatnagar:1954zz,Welander,Marle,Anderson:1974nyl}. One notable approach is the Anderson and Witting relaxation time approximation (RTA)~\cite{Anderson:1974nyl} of the collision term, where the collisional aspects of the medium can be quantified with a relaxation time parameter. Several efforts have been devoted in the formulation of relativistic dissipative hydrodynamics and in the estimation of transport coefficients within the RTA\footnote{Throughout the article, RTA refers to the relaxation time approximation introduced by Anderson and Witting, where the relaxation time is independent of particle energy or momentum.}~\cite{Strickland:2017kux, Denicol:2014tha, Jaiswal:2014isa, Kurian:2018dbn, Bhadury:2020ivo,Naik:2022pyk,Bhadury:2019xdf}. Notably, in the usual formulation of dissipative hydrodynamics with the RTA, it is assumed that the relaxation time is independent of particle momentum. However, in a realistic medium, the timescale of collisions is typically influenced by the specifics of microscopic interactions. Recently, there have been significant developments in modifying the standard RTA to address its fundamental issues, ensuring consistency, and preserving the basic properties of the linearized Boltzmann collision kernel~\cite{Rocha:2021zcw, Dash:2021ibx, Dash:2023ppc,Mitra:2020gdk,Calzetta:2024bpp}. In Refs.~\cite{Dash:2021ibx, Dash:2023ppc}, the authors established an Extended Relaxation Time Approximation (ERTA) framework for the systematic derivation of hydrodynamic equations from the Boltzmann equation. In this setup, a relaxation time dependent on particle momentum is introduced by extending the standard Anderson-Witting formulation. These recent developments regarding the collision kernel have not yet been considered in the formulation of magnetohydrodynamics. Setting up dissipative magnetohydrodynamics and transport coefficients of the magnetized QCD medium within the ERTA is an interesting task, which serves as the motivation for the current study.

In the present work, we derived the non-resistive, relativistic, second-order magnetohydrodynamic equations for shear viscous evolution with the momentum-dependent relaxation time for a conformal system with conserved charges. To that end, we generalized the ERTA approach to a magnetized medium. Utilizing the ERTA framework for the first time in a magnetized medium, we studied the magnetic field-dependent shear coefficients of the QCD matter. We solved the ERTA Boltzmann equation by using Chapman-Enskog-like gradient expansion. In the Navier-Stokes limit, we demonstrated that the anisotropic shear viscous coefficients of the magnetized QGP undergo significant modifications with the new formulation, as compared to those obtained within the conventional RTA framework. We have also analyzed the coupling of the shear stress tensor evolution to that of the dissipative charge current. 

The manuscript is organized as follows: we analyzed the first and second-order shear evolution for a conformal, charge conserved fluid with energy-dependent relaxation time in Section~\ref{sec1}. In Section~\ref{sec2}, we set up the first and second-order viscous magneto-hydrodynamic evolution and the associated transport coefficients for the non-resistive fluid within the ERTA followed by the discussions. Finally, we summarise the work with an outlook in Section~\ref{sec3}.

{{\bf Notations and Conventions:}}  Throughout the analysis, we work with natural units $c=k_B=\hbar=1$, where $c$ is the velocity of light in vacuum, $k_B$ represents the Boltzmann constant, and $\hbar$ denotes the reduced Planck's constant. We consider the Minkowskian metric tensor $g_{\mu\nu}={\text {diag}}\,(1,-1,-1,-1)$. Hydrodynamic fluid four-velocity is denoted by $u_\mu$ which satisfies $u^\mu u_\mu=1$ and has the form $u^{\mu}=(1,0,0,0)$ in its local rest frame. The projection operator that is orthogonal to $u^\mu$ is defined as $\Delta^{\mu\nu} = g^{\mu\nu} - u^\mu u^\nu$. The quantity $\Delta^{\mu\nu}_{\alpha\beta}\equiv\frac{1}{2}(\Delta^\mu_\alpha\Delta^\nu_\beta +\Delta^\mu_\beta\Delta^\nu_\alpha)-\frac{1}{3}\Delta^{\mu\nu}\Delta_{\alpha\beta}$ represents the rank-four traceless symmetric projection operator orthogonal to $u^\alpha$ and $\Delta^{\alpha\beta}$.

\section{Shear stress evolution with ERTA, $\mu\neq 0$,  $B=0$}\label{sec1}
The energy-momentum tensor and the net-quark current of a conformal system (system of massless quarks and gluons) with conserved charges can be expressed in terms of the single particle phase-space distribution function as,
\begin{align}\label{1.1}
&T^{\mu\nu}=\int{dP\,p^{\mu}p^{\nu}(f+\bar{f})}=\varepsilon u^{\mu} u^{\nu} - P \Delta^{\mu\nu} + \pi^{\mu\nu},\\ 
& N^{\mu}=\int{dP\,p^{\mu}(f-\bar{f})}=n\,u^{\mu}+ n^{\mu},\label{1.2}
\end{align}
where $dP=d_g\,\frac{d^3\mid{\bf {p}}\mid}{(2\pi)^3 p_0}$ is the phase space factor with $d_g$ as the degeneracy of the particle species. The quantities $\varepsilon$, $P$, and $n$ denote the energy density, pressure, and the net quark number of the medium. The tensor decompositions will be modified in the presence of a magnetic field. Here, $\pi^{\mu\nu}$ and $n^\mu$ are the dissipative quantities, namely shear stress tensor and particle diffusion current, respectively, and the hydrodynamic fluid velocity $u^\mu$ is defined in the Landau frame. It is important to emphasize that the  bulk viscous pressure vanishes in a conformal system (massless case). The fundamental evolution equations of $\varepsilon$,  $n$, and $u^\mu$ can be obtained from projections of energy-momentum conservation, $\partial_{\mu}T^{\mu\nu}=0$, along and orthogonal to $u^{\mu}$, and from the particle four-current conservation, $\partial_{\mu}N^{\mu}=0$, as follows,
\begin{align}\label{2.3}
\dot{\varepsilon}+(\varepsilon+P)\theta-\pi^{\mu\nu}\sigma_{\mu\nu}&=0,\\
(\epsilon+P)\dot{u}^\mu - \nabla^\mu P + \Delta^\mu_\nu \partial_\gamma \pi^{\gamma\nu}&= 0,\\
\dot{n}+n\theta+\partial_{\mu}n^{\mu}&=0,\label{2.5}
\end{align}
where $\dot{A}=u^\mu \partial_{\mu}A$ denotes the co-moving derivative, $\nabla^{\mu}=\Delta^{\mu\nu}\partial_\nu$ is the spacial derivative,  $\theta=\partial_{\mu}u^{\mu}$ is the expansion parameter, and $\sigma^{\mu\nu}=\Delta^{\mu\nu}_{\alpha\beta}\nabla^{\alpha}u^{\beta}$ is the velocity stress tensor. Here, we consider the particle equilibrium distribution function as $f_0=e^{-\beta(u \cdot p)+\alpha}$ where $\beta=\frac{1}{T}$ and $\alpha=\frac{\mu}{T}$ with $\mu$ as the chemical potential of the particle. For antiparticles, $f_0\rightarrow \bar{f}_0$ with $\alpha\rightarrow -\alpha$. We can define thermodynamic quantities from Eq.~(\ref{1.1}) and Eq.~(\ref{1.2}) as,
\begin{align}\label{2.6}
&\varepsilon=u_\mu u_\nu \int{dP\,p^{\mu}p^{\nu}(f_0+\bar{f}_0)}=6 d_g \frac{\cosh \alpha}{\pi^2 \beta^4},\\
& P=-\frac{\Delta_{\mu_\nu}}{3}\int{dP\,p^{\mu}p^{\nu}(f_0+\bar{f}_0)}=2 d_g \frac{\cosh \alpha}{\pi^2 \beta^4},\\
& n=u_\mu\int{dP\,p^{\mu}(f_0-\bar{f}_0)}=2 d_g \frac{\sinh \alpha}{\pi^2 \beta^3}. \label{2.8}
\end{align}
The expressions for the derivatives of $\beta$ and $\alpha$ can be obtained by substituting Eqs.~(\ref{2.6})-(\ref{2.8}) into Eqs.~(\ref{2.3})-(\ref{2.5}) as,
\begin{align}\label{ma2.9}
         \dot{\alpha} &= -A_n \partial_\mu n^\mu - A_\Pi \pi^{\mu\nu}\sigma_{\mu\nu}, \\
         \dot{\beta} &= \frac{\beta \theta}{3} - D_n \partial_\mu n^\mu - D_\Pi \pi^{\mu \nu} \sigma_{\mu\nu}, \\
         \nabla^\alpha \beta &= - \beta \dot{u}^\alpha + \frac{n}{\epsilon+P} \nabla^\alpha \alpha - \frac{\beta}{\epsilon+P} \Delta^\alpha_\nu \partial_\mu \pi^{\mu\nu},\label{ma2.11}
\end{align}
 where the coefficients $A_n$, $A_\Pi$, $D_n$ and $D_\Pi$ are functions of temperature inverse, $\beta$ and chemical potential $\alpha$, and hence a function of space-time $(x,t)$:
 \begin{align}\label{m2.12}
     &A_n = \frac{\beta \epsilon /3}{\left( \frac{\epsilon^2 \beta^2}{9} - \frac{3n^2}{4} \right)}, && A_\Pi = \frac{\beta n/4}{\left( \frac{\epsilon^2 \beta^2}{9} - \frac{3n^2}{4} \right)}, \\
     &D_n = \frac{3 n \beta}{\left( \frac{4 \epsilon^2 \beta^2}{3} - 9 n^2 \right)}, &&    D_\Pi = \frac{\epsilon \beta^3}{3 \left( \frac{4 \epsilon^2 \beta^2}{3} - 9 n^2 \right)}.\label{m2.13}
\end{align}    
For a system near to local thermodynamic equilibrium, the particle non-equilibrium phase-space distribution function can be expressed as $f=f_0+\delta f$ with $\mid{\delta f}\mid\ll {f_0}$ (for antiparticles $\bar{f}=\bar{f}_0+\delta \bar{f}$). Using Eq.~(\ref{1.1}) and Eq.~(\ref{1.2}), the shear stress tensor and the particle diffusion current can be defined in terms of the non-equilibrium part of the distribution function as,
\begin{align}\label{2.9}
&\pi^{\mu\nu}=\Delta^{\mu\nu}_{\alpha\beta}\int{dP\,p^{\alpha}p^{\beta}(\delta f+\delta \bar{f})},\\ 
& n^{\mu}=\Delta^\mu_\nu\int{dP\,p^{\nu}(\delta f-\delta \bar{f})}.\label{2.10}
\end{align}
The derivation of these dissipative quantities requires the knowledge of $\
\delta f$ and $\delta \bar{f}$ and can be obtained by solving the Boltzmann equation. We obtain the first and second-order corrections to the local equilibrium distribution  within the ERTA framework in the next section.
\subsection{Solving Boltzmann equation with ERTA}\label{2.1}
The relativistic transport equation characterizes the evolution of the phase-space distribution function away from equilibrium. In the absence of electromagnetic fields, the Boltzmann equation (for particles) takes the form as follows,
\begin{align}
p^\mu \partial_\mu f= C(f),   
\end{align}
where $C(f)$ is the collision kernel. The conventional way to quantify the collision terms is to adopt the RTA framework introduced by Anderson and Witting as $C(f)=-\frac{(u\cdot p)}{\tau_R(x)}$ with $\tau_R(x)$ as the relaxation time~\cite{Anderson:1974nyl}. However, the above approximation assumes the relaxation time to be independent of particle energy. As the timescale of collisions depends on the microscopic interactions in the medium, we consider momentum-dependent relaxation time in the present analysis. When the $\tau_R$ becomes dependent on particle energy, it is seen that the conventional RTA is incompatible with conservation equations for the energy-momentum tensor and net-charge current, even in the Landau frame. Hence, we work with the recently developed ERTA framework~\cite{Dash:2021ibx} that takes into account the energy-dependent relaxation time $\tau_R(x, p)$ by ensuring the satisfaction of conservation equations. The Boltzmann equation within the ERTA has the form,
\begin{align}\label{2.12}
p^\mu \partial_\mu f=-\frac{(u\cdot p)}{\tau_R(x, p)} \Big(f - f^*_0 \Big).   
\end{align}
Here, $f^*_0=e^{-\beta^*(u^*\cdot p)+\alpha^*}$ with $\beta^*=\frac{1}{T^*}$ and $\alpha^*=\frac{\mu^*}{T^*}$. The four-vector $u^*_\mu$  does not necessarily have to correspond to the hydrodynamic four-velocity $u^\mu$. In the local rest frame of $u^*_\mu$, which can be referred to as the thermodynamic frame, the distribution function is reduced to the Maxwell-Boltzmann form with $T^*$ and $\mu^*$ as the temperature and chemical potential, respectively. The interpretation and distinctions between the thermodynamic frame and the hydrodynamic frame are discussed in detail in Ref.~\cite{Dash:2021ibx}. Non-equilibrium corrections to the distribution function are obtained by employing an iterative Chapman-Enskog-like solution of the ERTA Boltzmann equation as, 
\begin{align}
     f=f_0+ \delta f_{(1)}+ \delta f_{(2)} + \cdots , 
\end{align}
where $\delta f_{(i)}$ with $(i=1, 2, 3, ...)$ is the  gradient correction of the distribution function to the $i^{\text{th}}$ order. Note that the expansion is carried out around the local hydrodynamic equilibrium as our interest lies in deriving hydrodynamic equations.
\subsection{First and second-order shear viscous evolution at finite $\mu$}
\subsubsection{First-order evolution}
We briefly review the derivation of the first-order evolution equation of shear tensor and number diffusion for a conformal system. The first-order correction to the particle distribution functions can be obtained from the ERTA Boltzmann equation as defined in Eq.~(\ref{2.12}) as follows,
\begin{align}\label{2.14}
    \delta f_{(1)} = -\frac{\tau_R}{(u \cdot p)}p^\mu \partial_\mu f_0 + \delta f^*_{(1)},
\end{align}
where $\delta f^*_{(1)}=f^*_0-f_0$ arises due to the difference between the definition of thermodynamic and hydrodynamic frame variables. Defining  $T^*=T+\delta T$, $\mu^*=\mu+\delta \mu$, and $u_\mu^*=u_\mu+\delta u_\mu$, we can define $\delta f^*_{(1)}$ from the Taylor expansion of $f^*_0$ about $T$, $\mu$ and $u_\mu$ as,
\begin{align}\label{2.15}
  &\delta f^*_{(1)} = \left[ -\frac{(\delta u \cdot p)}{T} + \frac{(u \cdot p - \mu)}{T^2} \delta T + \frac{\delta \mu}{T}\right] f_0.  
\end{align}
Employing the form of the hydrodynamic equilibrium distribution function and Eq.~(\ref{2.15}) in Eq.~(\ref{2.14}), we can write the first-order correction to the particle and antiparticle distribution functions as,
\begin{align}\label{2.16}
    &\delta f_{(1)} = \tau_R \left[\left(\frac{n}{\epsilon+P} - \frac{1}{(u \cdot p)}\right) p^\mu \nabla_\mu \alpha + \frac{\beta p^\mu p^\alpha \sigma_{\mu \alpha}}{(u \cdot p)} \right] f_0 + \left[ -\frac{(\delta u \cdot p)}{T} + \frac{(u \cdot p - \mu)}{T^2} \delta T + \frac{\delta \mu}{T}\right] f_0,\\
    &\delta \bar{f}_{(1)}= \tau_R \left[\left(\frac{n}{\epsilon+P} + \frac{1}{(u \cdot p)}\right) p^\mu \nabla_\mu \alpha + \frac{\beta p^\mu p^\alpha \sigma_{\mu \alpha}}{(u \cdot p)} \right] \bar{f}_0+\left[ -\frac{(\delta u \cdot p)}{T} + \frac{(u \cdot p + \mu)}{T^2} \delta T - \frac{\delta \mu}{T}\right] \bar{f}_0.\label{2.17}
\end{align}    
The quantities $\delta T$, $\delta \mu$, and $\delta u_\mu$ can be defined by imposing the Landau frame condition $u_\nu T^{\mu\nu}=\varepsilon u^\mu$ and the matching conditions, $\varepsilon=\varepsilon_0$ and $n=n_0$. In the most general case, these quantities take the form as~\cite{Dash:2021ibx},
\begin{align}\label{2.18}
  &\delta u^\mu=C\beta \nabla^\mu \alpha, && \delta \mu=\Bar{C} \theta, && \delta T= \mathfrak{C} \theta,
\end{align}
where $C$, $\Bar{C}$, and $\mathfrak{C}$ are the dimensionless variables.  
It can be shown that for a conformal system with conserved charges, the variables $\Bar{C}$ and $\mathfrak{C}$ vanishes and $C$ takes the form as,
\begin{align}\label{2.18re}
    C=\frac{1}{\beta^2 I_{31}^+}\left( \frac{n}{\epsilon+P}K_{31}^+ - K_{21}^- \right).
\end{align}
Here, $I^{(r)\pm}_{nq}$ and $K^{(r)\pm}_{nq}$ are the thermodynamic integrals as defined in Appendix~\ref{AA}. Employing Eqs.~(\ref{2.16})-(\ref{2.18}) in Eq.~(\ref{2.9}) and Eq.~(\ref{2.10}), we obtain the expression of shear tensor and number diffusion current as,
\begin{align}\label{re25}
    &\pi^{\mu\nu} = 2 \eta_0 \sigma^{\mu\nu}, \hspace{1cm} \text{with}\,\,\,\,\, \eta_0 = \frac{K_{32}^+}{T} \\
    &n^\mu = \kappa \nabla^\mu \alpha,  \hspace{1.2cm} \text{with}\,\,\,\,\, \kappa = \left( -C\beta^2 I^-_{21}+ \frac{n}{\epsilon+P}K^-_{21} - K^+_{11} \right).\label{re26}
\end{align}
{Before proceeding to the derivation of the second-order evolution equation for the shear stress tensor, let us discuss the significance of the quantities,  $\mu^*=\mu+\delta \mu$, $T^*=T+\delta T$, and $u_\mu^*=u_\mu + \delta u_\mu$ in the ERTA framework. The conventional formulation of hydrodynamics with RTA (momentum independent $\tau_R$) works within the Landau frame to uphold macroscopic conservation laws. Simply replacing the conventional thermal relaxation time with momentum-dependent relaxation time can result in the violation of microscopic conservation laws. This has generated significant attention towards achieving a consistent approximation of the collision term. A recent study~\cite{Rocha:2021zcw} offers a notable advancement in this direction by introducing counterterms in the form of projectors to restore the conservation laws in the presence of momentum-dependent relaxation time, regardless of the choice of hydrodynamic frames.
The ERTA framework presents a distinct solution to this problem.
By defining the equilibrium distribution function within the RTA approximation in the "thermodynamic frame" (characterized by the parameters $\mu^*$, $T^*$, and $u_\mu^*$), one can construct consistent order-by-order hydrodynamics such that conservation laws are satisfied at every stage of the gradient expansion. Since our focus is on the derivation of hydrodynamic equations, the expansion of the non-equilibrium distribution function is done about the hydrodynamic equilibrium. However, the non-equilibrium part of the distribution function gets an additional contribution in the current framework $\delta f^*$ that arises from the difference between the definition of thermodynamic and hydrodynamic frame variables ($i.e.$, in terms of $\mu^*-\mu=\delta \mu$, $T^*-T=\delta T$, and $u_\mu^*-u_\mu=\delta u_\mu$). The counter terms due to the momentum dependence of the relaxation time are thus absorbed in the definition of $\delta\mu$, $\delta T$, and $\delta u_\mu$.  In Appendix~\ref{AD}, we have presented a detailed derivation of the conservation equations within the ERTA setup and verified that the definition of $\delta \mu$, $\delta T$, and $\delta u_\mu$ encompasses the necessary counter terms arising from the momentum dependence of the relaxation time, thus ensuring the satisfaction of the conservation equations.}

\subsubsection{Second-order evolution}
In Ref.~\cite{Jaiswal:2015mxa}, the authors have explored second-order hydrodynamics at a finite chemical potential within the conventional RTA framework. One important observation from the study was that, within the RTA, the second-order evolution equations for $\pi^{\mu\nu}$ and $n^\mu$ can be decoupled for a system of massless quarks and gluons. In this subsection, we re-examine the second-order evolution of shear viscosity at finite $\mu$ using the ERTA framework. We start by expressing the distribution function until second-order in gradient expansion as $f=f_0+ \Delta f$ with $\Delta f\equiv\delta f_{(1)}+ \delta f_{(2)}$. Employing the Chapman-Enskog-like expansion and ERTA Boltzmann equation, we have
\begin{align}\label{2.24}
    \Delta f = -\frac{\tau_R}{(u \cdot p)} p^\mu\partial_\mu f_0 - \frac{\tau_R}{(u \cdot p)} p^\mu \partial_\mu \Big[-\frac{\tau_R}{(u \cdot p)}p^\mu \partial_\mu f_0 + \delta f^*_{(1)}\Big] + \Delta f_{(2)}^*.
\end{align}
Employing the definitions Eqs.~(\ref{ma2.9})-(\ref{ma2.11}) and keeping all terms till second order, the first term of Eq.~(\ref{2.24}) can be simplified as, 
\begin{align}\label{2.25}
      -\frac{\tau_R}{(u \cdot p)}&p^\mu \partial_\mu f_0 = \tau_R \Bigg[\left(\frac{n}{\epsilon+P} - \frac{1}{(u \cdot p)}\right) p^\mu \nabla_\mu \alpha + \frac{\beta p^\mu p^\alpha \sigma_{\mu \alpha}}{(u \cdot p)} -\frac{\beta}{\epsilon+P}\left\{p^l \nabla_k \pi^{k}_l - p^l \pi_{kl}\dot{u}^k\right\}\nonumber \\
      &+\left(A_n -D_n(u\cdot p)\right)\partial_\mu n^\mu +\left\{A_\Pi -\left(D_\Pi+\frac{\beta}{\epsilon+P}\right) (u \cdot p))\right\} \pi^{\alpha\beta}\sigma_{\alpha\beta} \Bigg] f_0.
\end{align}
Notably, for massless system at $\mu=0$ case, $\frac{\tau_R}{(u \cdot p)} p^\mu \partial_\mu  \delta f^*_{(1)}=\mathcal{O}(\partial^3)$ and will not contribute to the second-order evolution equation. However, this term will have a non-vanishing contribution at finite $\mu$. The second term of Eq.~(\ref{2.24}) in the case of massless charge conserved system reduces as,
\begin{align}
- \frac{\tau_R}{(u \cdot p)} p^\mu \partial_\mu &\Big[-\frac{\tau_R}{(u \cdot p)}p^\mu \partial_\mu f_0 + \delta f^*_{(1)}\Big]=-\frac{\tau_R}{(u \cdot p)} p^k\partial_k \Bigg[ \Bigg( -C\frac{p^l \nabla_l \alpha}{T^2} + \tau_R  \left(\frac{n}{\epsilon+P} - \frac{1}{(u \cdot p)}\right)p^l \nabla_l \alpha+ \tau_R \beta \frac{p^lp^m \sigma_{lm}}{(u \cdot p)} \Bigg)f_0 \Bigg].
\end{align}
Imposing the Landau frame condition and matching condition as in first-order, one can obtain the last term of Eq.~(\ref{2.24}) as, 
\begin{align}
     &\Delta f^*_{(2)} = \left[ -\frac{(\delta u_{(2)} \cdot p)}{T} + \frac{(u \cdot p - \mu)}{T^2} \delta T_{(2)} + \frac{\delta \mu_{(2)}}{T}\right] f_0.
\end{align}
However, it is observed that the contribution from $\Delta f^*_{(2)}$ vanishes in the context of shear viscous evolution\footnote{It is seen that $\Delta^{\mu\nu}_{\alpha\beta}\int \mathrm{dP} p^\alpha p^\beta \Big(\Delta f^*_{(2)}+\Delta \bar{f}^*_{(2)}\Big) =-\Delta^{\mu\nu}_{\alpha\beta} I^{\alpha \beta \gamma}_+ \frac{\Delta u_{\gamma{(2)}}}{T} +\Delta^{\mu\nu}_{\alpha\beta} I^{\alpha \beta \gamma}_+ u_\gamma \frac{\Delta T_{(2)}}{T^2}- \Delta^{\mu\nu}_{\alpha\beta} I^{\alpha \beta}_- \mu \frac{\Delta T_{(2)}}{T^2} + \Delta^{\mu\nu}_{\alpha\beta} I^{\alpha \beta}_- \frac{\Delta \mu_{(2)}}{T}=0$ due to the properties of our-rank traceless symmetric projection operator. The decomposition of $I^{\alpha \beta \gamma}$ is given in Eq.~(\ref{A7}). }. It is important to emphasize that the evolution of bulk viscous pressure and number diffusion will depend on the term $\Delta f^*_{(2)}$. Substituting Eq.~(\ref{2.24}) in Eq.~(\ref{2.9}), we obtain the second-order evolution equations for shear tensor as,
\begin{align}\label{2.28}
        \dot{\pi}^{\langle\mu\nu\rangle} + \frac{\pi^{\mu\nu}}{\tau_\pi} =  2\beta_\pi \sigma^{\mu\nu} - \delta_{\pi\pi} \pi^{\mu\nu}\theta +2\pi_\gamma^{\langle\mu}\omega^{\nu\rangle\gamma} - \tau_{\pi\pi}\pi_\gamma^{\langle\mu}\sigma^{\nu\rangle\gamma}-\tau_{\pi n} n^{\langle \mu}\dot{u}^{\nu\rangle}+\lambda_{\pi n} n^{\langle \mu}\nabla^{\nu\rangle}\alpha+ l_{\pi n} \nabla^{\langle\mu}n^{\nu\rangle} .
\end{align}
We have defined $\tau_{\pi}=\frac{L^+_{32}}{K^+_{32}}$ such that $\tau_{\pi} \rightarrow \tau_R$ in the limit where the relaxation time is independent of particle momentum. The transport coefficients appearing in the shear viscous evolution in Eq.~\eqref{2.28} are obtained as,
\begin{align}\label{mm2.32}
    \beta_\pi &= \frac{\eta_0}{\tau_\pi}= \frac{(K^+_{32})^2}{T L^+_{32}},\\
    \delta_{\pi\pi} &=\frac{4}{3},\\
    \tau_{\pi\pi} &= \frac{2}{7} \frac{L^+_{42}}{T L _{32}^+},\\
    \tau_{nn} &= 2\left( C\beta^2 K^+_{32}-\Tilde{\chi} L^+_{32}+L^-_{22}\right), \\
    l_{\pi n} &= \frac{2}{\kappa \tau_\pi}\left( C\beta^2 K^+_{32}-\Tilde{\chi} L^+_{32}+L^-_{22}\right),\\
    \tau_{\pi n} &= \frac{2}{\kappa \tau_\pi}\Bigg[ \Tilde{\chi} M_{42}^+ - M^-_{32} + L_{22}^- +  K_{32}^+ \left( 2\beta^2 C - \beta^3 \frac{\partial C}{\partial \beta} \right) - \beta L_{32}^+ \frac{\partial \Tilde{\chi}}{\partial \beta} -\Tilde{\chi} N_{32}^+ \beta + N_{22}^- \beta -\frac{\tau_{nn}\beta}{2\kappa}\frac{\partial \kappa}{\partial \beta}\Bigg],\\
    \lambda_{\pi n} &= \frac{1}{\kappa \tau_\pi}\Bigg[K_{32}^+ \Bigg\{4C\beta \Tilde{\chi}  + 2 \beta^2 \left( \frac{\partial C}{\partial \alpha} + \Tilde{\chi} \frac{\partial C}{\partial \beta} \right)\Bigg\} - 2 \Tilde{\chi}^2 N_{32}^+ + 2 \Tilde{\chi} N_{22}^- -2L_{32}^+\left( \frac{\partial \Tilde{\chi}}{\partial \alpha} + \Tilde{\chi} \frac{\partial \Tilde{\chi}}{\partial \beta} \right)\nonumber\\
    & + 2 C \beta^2 K_{32}^- - 2 C \beta^2\Tilde{\chi} K_{42}^+ - 4\Tilde{\chi}  L_{32}^- + 2 L_{22}^+ + 2\Tilde{\chi}^2 L_{42}^+-\frac{\tau_{nn}}{\kappa}\left(\frac{\partial \kappa}{\partial \alpha} + \Tilde{\chi} \frac{\partial \kappa}{\partial \beta}\right)\Bigg], \label{mm2.37}
\end{align}
with $\Tilde{\chi} = \frac{n}{\epsilon+P}$.
\begin{figure}
\includegraphics[scale=0.430]{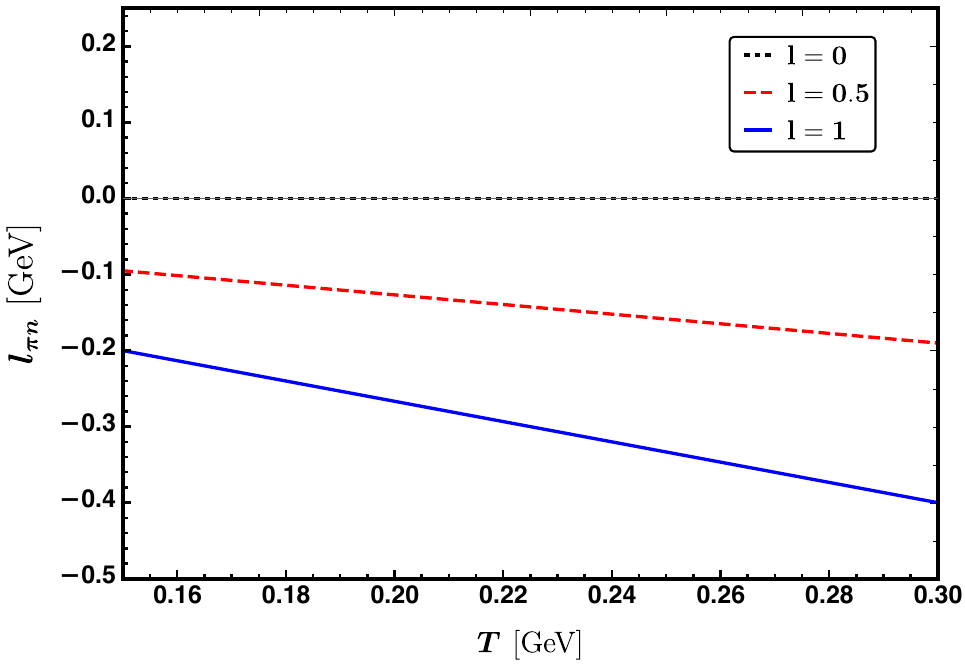}
\hspace{-.25 cm }
\includegraphics[scale=0.430]{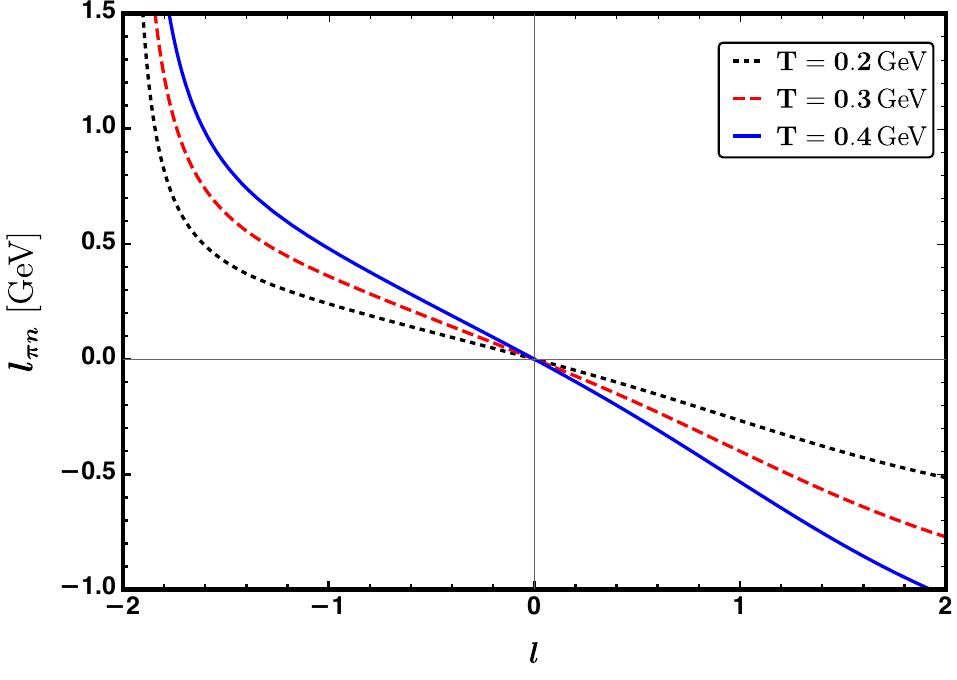}
\includegraphics[scale=0.430]{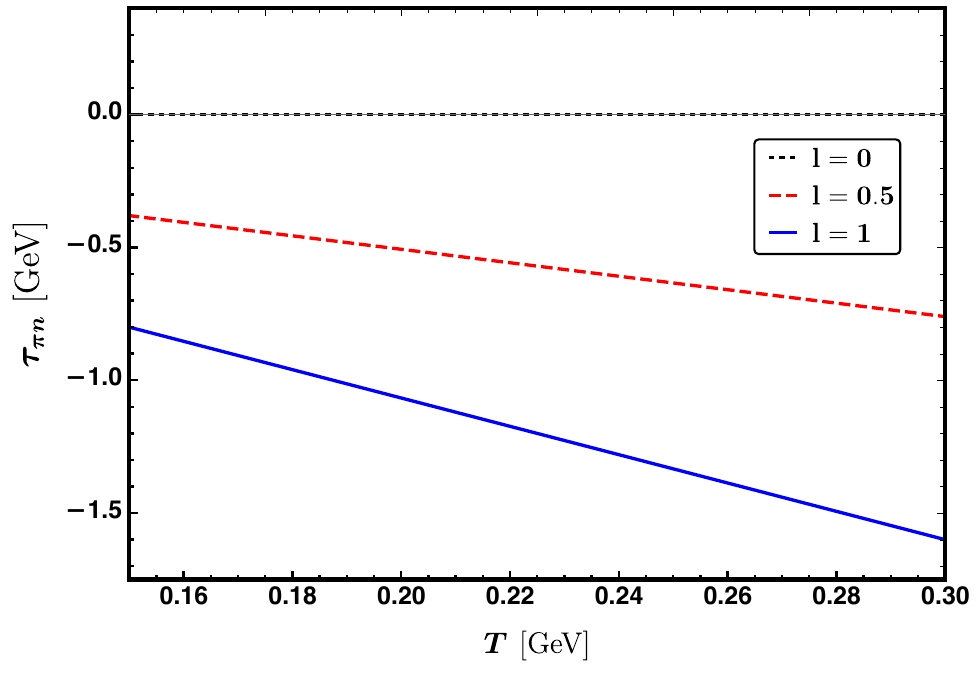}
\includegraphics[scale=0.430]{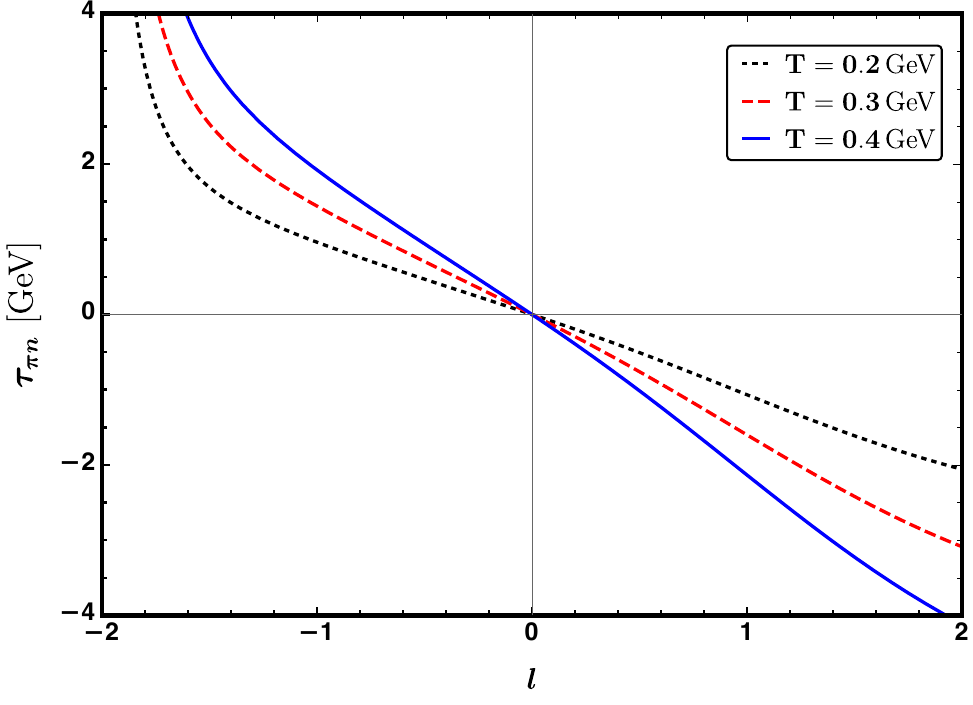}
\includegraphics[scale=0.42]{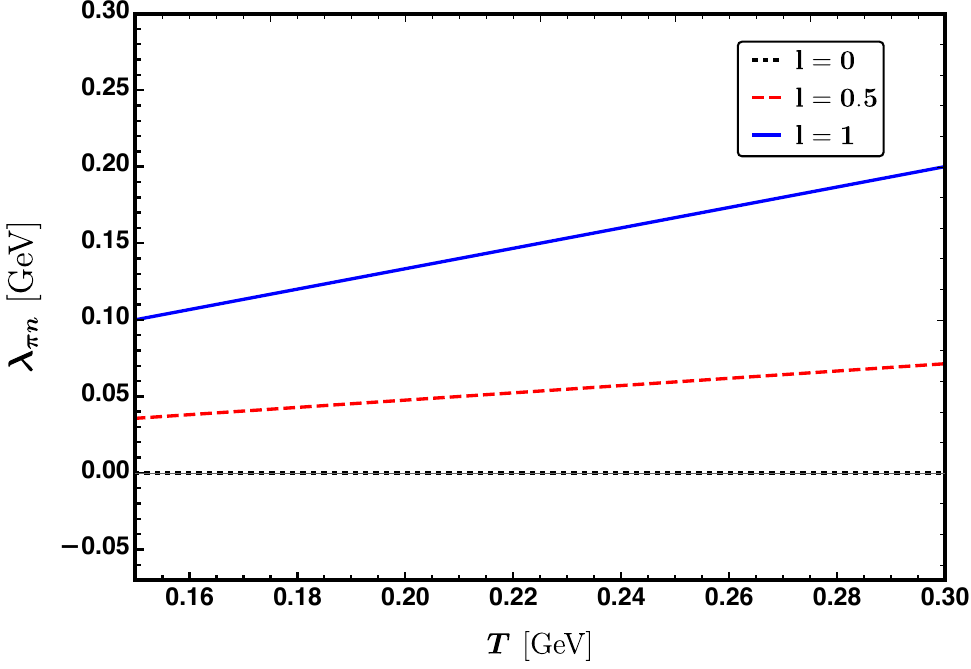}
\hspace{-.3 cm }
\includegraphics[scale=0.45]{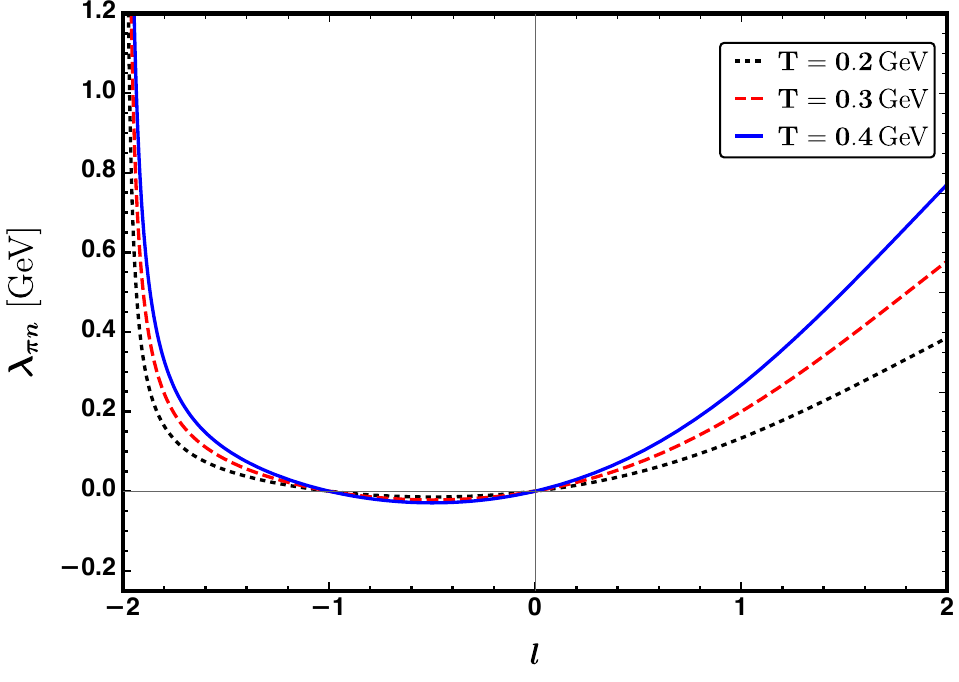}
\caption{\small Impact of momentum-dependent relaxation time on the temperature dependence of  $l_{\pi n}$,  $\tau_{\pi n}$ and  $\lambda_{\pi n}$. The coefficients are also plotted as a function of $\ell$ at different temperatures. The ERTA results are compared with those from RTA estimations. Here, $\ell=0$ corresponds to the usual RTA approximation. }
\label{figure1}
\end{figure}
The thermodynamic integrals, $L_{nq}$, $M_{nq}$, and $N_{nq}$ are defined in Appendix~\ref{AA}. The Eq.~\eqref{2.28} denotes the second-order evolution equation of the shear stress tensor for a charge conserved conformal fluid with the new momentum-dependent ERTA framework. Shear viscosity with energy-dependent relaxation time in the absence of dissipative charge current in the QCD medium has been analyzed recently~\cite{Dash:2023ppc}. In Fig.\ref{figure1}, we depicted the temperature behaviour of new coefficients $\lambda_{\pi n}$, $l_{\pi n}$, and $\tau_{\pi n}$ that couples shear tensor with charge current for a system of massless quarks and gluons at a finite chemical potential with the ERTA approach. For quantitative estimation, we use a power law parametrization for the relaxation time as~\cite{Dusling:2009df, Chakraborty:2010fr, Dusling:2011fd, Kurkela:2017xis},
\begin{align}\label{m2.38}
    \tau_R(x,p)=\tau_0(x) \Big(\frac{u\cdot p}{T}\Big)^\ell, && \text{where,}\,\,\, \tau_0(x)=\bar{\kappa}/T.
\end{align}
Here, $\tau_0(x)$ is the momentum-independent part, $\bar{\kappa}$ being a dimensionless constant, and $\ell$ is an arbitrary parameter that determines the order of momentum dependence of the relaxation time. The parameter $\tau_0$ is of a time scale proportional to the mean free path and hence we can use $\tau_0(x)$ to define our Knudsen number which is $Kn=\tau_0\partial$~\cite{Denicol:2014vaa}. In the limit $\ell=0$, we have $\tau_R=\tau_0(x)$ which describes the momentum-independent conventional RTA framework. {Taking $\bar{\kappa}$ to be a dimensionless constant here, in the massless, Maxwell-Boltzmann case, we can express the above transport coefficients only as a function of the momentum dependence parameter, $\ell$, chemical potential $\alpha$ and temperature $T$ as follows:
\begin{align}
    \tau_\pi &= \frac{\bar{\kappa} \Gamma(5+2\ell)}{T\Gamma(5+\ell)}, \quad \ell>-\frac{5}{2} \\
    \eta_0 &=\frac{e^\alpha d_g \bar{\kappa} \Gamma(5+\ell)}{30 \pi^2 \beta^3}, \quad \ell> -5\\
    \tau_{\pi\pi} &= \frac{2 \Gamma(6+2\ell)}{7\Gamma(5+\ell)}, \quad \ell>-\frac{5}{2} \\
    l_{\pi n} &= \frac{T \ell \Gamma(5+\ell) \left\{ \Gamma(5+\ell)\Gamma(4+\ell) - 48 \Gamma(4+2\ell) \right\}}{15 (\ell^2 - \ell +4) \Gamma(3+\ell)\Gamma(5+2\ell)}, \quad \ell>-2 \\
    \tau_{\pi n} &= \frac{4T \ell \Gamma(5+\ell) \left\{ \Gamma(5+\ell)\Gamma(4+\ell) - 48 \Gamma(4+2\ell) \right\}}{15 (\ell^2 - \ell +4) \Gamma(3+\ell)\Gamma(5+2\ell)}, \quad \ell>-2\\
    \lambda_{\pi n} &= \frac{\ell(\ell +1)T \Gamma(5+\ell) \left\{-\Gamma(4+\ell)\Gamma(5+\ell)+48\Gamma(4+2\ell)\right\}}{60(\ell^2 - \ell +4)\Gamma(3+\ell)\Gamma(5+2\ell)}, \quad \ell>-2
\end{align}
}
In a prior study, detailed in Ref.~\cite{Jaiswal:2015mxa}, the authors have demonstrated that the second-order evolution equations for the shear stress tensor and the charge current are decoupled for a system of massless quarks and gluons at finite chemical potential within the conventional RTA. In contrast to this, with the present analysis within the ERTA, we observe that there exists a coupling with the evolution of shear tensor and particle diffusion in the medium due to the momentum dependence of the thermal relaxation time. In Eq.~\eqref{2.28} the terms  $n^{\langle \mu}\dot{u}^{\nu\rangle}$, $n^{\langle \mu}\nabla^{\nu\rangle}\alpha$, and $\nabla^{\langle\mu}n^{\nu\rangle}$ couples evolution of shear tensor with particle current. We observe that Eq.~\eqref{2.28} reduces to the result of \cite{Jaiswal:2015mxa}, when we switch off the momentum dependence of the relaxation time as the coefficients $\lambda_{\pi n}$, $l_{\pi n}$, and $\tau_{\pi n}$ vanishes for a system of massless particles at finite $\mu$ in the limit of constant $\tau_R$. It is seen that the momentum dependence of relaxation time has a significant effect on the transport coefficients. 
All thermodynamic integrals are modified in the ERTA approach, and notably,  a new thermodynamic integral $M^{(m)\pm}_{nq}$ as defined in Eq.~(\ref{A10}) arises solely due to momentum dependence of the relaxation time. In addition to that, the term $C$ (that arises from the matching condition) gives a non-vanishing contribution within the ERTA framework. We observed that $\lambda_{\pi n}$, $l_{\pi n}$, and $\tau_{\pi n}$ are sensitive to the value of $\ell$. The sign of $\ell$ also plays an important role in the estimation as it indicates how the particle energy influences the strength of interaction, $i.e$, a positive sign of $\ell$ indicates the decrease of interaction strength with the energy of the particle. Notably, the sign of the coefficients gets reversed with the change in the sign of $\ell$. It can be argued that range $0\le l\le 1$ is more relevant for QCD medium~\cite{Dusling:2009df}. The signs of the transport coefficients in the range $0\le l\le 1$ align with those found in the previous study for QCD medium~\cite{Denicol:2012es}.  
\subsection{Comparison  of the ERTA results with exact results from $\lambda \phi^4$ theory ($\ell=1$ case)}
Exact analytical solutions of the Boltzmann equation are difficult to obtain and have been estimated only for homogeneous, isotropic systems with a simpler interaction model. We follow Refs.~\cite{Denicol:2022bsq,Rocha:2023hts} where the  exact results of transport coefficients for a massless scalar theory are obtained. In the recent study~\cite{Denicol:2022bsq}, the authors have analytically extracted a full set of eigenvalues and eigenfunctions of the relativistic linearized Boltzmann collision operator with the interaction considered being $\lambda \phi^4$ (massless scalar theory). Further, it has been shown that in the appropriate limit, a linearised collision kernel leads to the recently formulated novel RTA proposed in Ref.~\cite{Rocha:2021zcw}.

{Considering the massless scalar $\lambda \phi^4$ theory at high temperatures, the linearised collision operator can be written as~\cite{Denicol:2022bsq},
\begin{equation}
    \hat{L}\phi_k = \frac{g}{2} \int dK'dPdP' f_{0k'}(2\pi)^5\delta^{(4)}(k+k'-p-p')(\phi_p + \phi_{p'} - \phi_{k} - \phi_{k'}).
\end{equation}
The eigenfunctions and eigenvalues of this operator were calculated exactly and they are given by~\cite{Denicol:2022bsq},
\begin{equation}
\hat{L} L_{n \mathbf{k}}^{(2 m+1)} k^{\left\langle\mu_1\right.} \ldots k^{\left.\mu_{\ell}\right\rangle}=-\frac{g \mathcal{M}}{2}\left[\frac{n+m-1}{n+m+1}+\delta_{\ell 0} \delta_{n 0}\right] L_{n \mathbf{k}}^{(2 m+1)} k^{\left\langle\mu_1\right.} \ldots k^{\left.\mu_{m}\right\rangle},
\end{equation}
where $L^{(2m+1)}_{n \mathbf{k}}$ are the associated Laguerre polynomials of degree $n$. With this, various transport coefficients have been calculated in first and second order in gradient expansion~\cite{Rocha:2023hts}. It has also been noted that if we expand $\phi_k$ on the basis of the irreducible tensors that constitute the eigenfunctions of $\hat{L}$, we recover the RTA with momentum-dependent relaxation time, $\tau_R$ along with some counter-terms. The form of $\tau_R$ can be read off from $\hat{L} \phi_k$ as,
\begin{equation}\label{43}
    \tau_R(p) = \frac{2(u \cdot p)}{g \mathcal{M}}.
\end{equation}
With $\mathcal{M} = e^\alpha/(2\pi^2 \beta^2) $ and $g=\lambda^2 / (32\pi)$ and $\tau_R$ takes the form as follows,
\begin{equation}\label{ns1}
    \tau_R = \frac{4\pi^2}{g e^\alpha T} \left(\frac{u \cdot p}{T}\right).
\end{equation}
Comparing Eq.~(\ref{m2.38}) and Eq.~(\ref{ns1}), in $\lambda \phi^4$, we identify $\bar{\kappa}$ in the  ERTA framework as $\bar{\kappa}= 4\pi^2/(g e^\alpha)$. With this, we can compare the results of the transport coefficients calculated in the ERTA for $\ell=1$ with those from $\lambda \phi^4$ theory. With this identification of $\bar{\kappa}$, when the mass is taken to be zero, Eq.~(\ref{mm2.32}) - Eq.~(\ref{mm2.37}) can be expressed as a function of only $T$ and the momentum-dependence parameter, $\ell$  for comparison with the $\lambda \phi^4$ theory:
\begin{align}
    \eta_0 &= \frac{2d_g \Gamma(5+\ell)}{15g \beta^3}, \quad \ell>-5 \\
    \tau_{\pi} &= \frac{4\pi^2}{g e^\alpha T} \frac{\Gamma(5+2\ell)}{\Gamma(5+\ell)}, \quad l>-\frac{5}{2} \\
    \tau_{\pi\pi} &= \frac{2}{7} \frac{\Gamma (6+2\ell)}{\Gamma(5+2\ell)}, \quad \ell> -\frac{5}{2}\\
    l_{\pi n} &= \frac{T \ell \Gamma(5+\ell) \left\{ \Gamma(5+\ell)\Gamma(4+\ell) - 48 \Gamma(4+2\ell) \right\}}{15 (\ell^2 - \ell +4) \Gamma(3+\ell)\Gamma(5+2\ell)}, \quad \ell>-2 \\
    \tau_{\pi n} &= \frac{4T \ell \Gamma(5+\ell) \left\{ \Gamma(5+\ell)\Gamma(4+\ell) - 48 \Gamma(4+2\ell) \right\}}{15 (\ell^2 - \ell +4) \Gamma(3+\ell)\Gamma(5+2\ell)}, \quad \ell>-2\\
    \lambda_{\pi n} &= \frac{T \ell(5+\ell ) \left\{ -(\ell +1)\Gamma(4+\ell)\Gamma(5+2\ell) - 48(\ell-3)\Gamma(4+2\ell) \right\}}{60(\ell^2 - \ell +4)\Gamma(3+\ell)\Gamma(5+2\ell)}, \quad \ell>-2
\end{align}
\vspace{1mm}
The comparison is summarised in Table~\ref{t1} for $\ell=1$ \footnote{{For comparison purpose, we have rearranged the second-order shear evolution equation of Ref.~\cite{Rocha:2023hts} as in the form of Eq.~(\ref{2.28})}}.
\begin{table}
    \begin{tabular}{|c|c|c|c|}
     \hline   
      Coefficients   & RTA results $(l=0)$ & ERTA results $(l=1)$ & $\lambda \phi^4$ results (exact) \\
      \hline
      $\tau_\pi$ & $\tau_c$ & $\frac{24 d_g}{g n_0 \beta^2}$ & $\frac{72}{g n_0 \beta^2}$\\
    \hline
      $\eta$ & $\frac{4P}{5}$ & $\frac{16d_g}{g\beta^3}$ &$\frac{48}{g\beta^3}$\\
      \hline
      $\kappa$ & $\frac{n_0}{12}$ & $\frac{d_g}{g\beta^2}$ & $\frac{3}{g\beta^2}$ \\
     \hline    
   $\delta_{\pi \pi}$ & $\frac{4}{3}$ & $\frac{4}{3}$ & $\frac{4}{3}$ \\
\hline
$\tau_{\pi\pi}$ & $\frac{10}{7}$ & $2$ & $2$ \\
\hline
$l_{\pi n}$ & $0$ & $-\frac{4}{3\beta}$ & $-\frac{4}{3\beta}$\\
\hline
$\tau_{\pi n}$ & $0$ & $-\frac{16}{3\beta}$ & $-\frac{16}{3\beta}$ \\
\hline
$\lambda_{\pi n}$ & $0$ & $-\frac{2}{\beta}$ & $\frac{5}{6\beta}$ \\
\hline
\end{tabular}
\caption{{Comparison of the ERTA coefficients with exact results from $\lambda \phi^4$ theory by taking $\bar{\kappa}$ as given in Eq.~(\ref{43}). The exact form of transport coefficients of $\lambda \phi^4$ theory are obtained from~\cite{Denicol:2022bsq, Rocha:2023hts}.} }
\label{t1}
\end{table}
We observe that $\tau_\pi \sim (T \; \text{exp} \; \alpha)^{-1}$ in the ERTA $(\ell=1)$ case, $i.e$, the relaxation time also has a dependence on the fugacity $\alpha$ of the gas in addition to the temperature, and this observation is consistent with the result in~\cite{Rocha:2023hts}. It is seen that the temperature dependence of first and most of the second order transport coefficients are consistent in comparison with those obtained from the $\lambda \phi^4$ theory. We also note that, $\eta$, $\kappa$ and $\tau_\pi$ values for ERTA $(\ell=1)$ depends on the degeneracy factor $d_g$ unlike the $\lambda \phi^4$ results.} 

{As for the last three second order transport coefficients, $l_{\pi n}$ and $\tau_{\pi n}$ matches with the exact results from $\lambda \phi^4$ theory. On the other hand $\lambda_{\pi n}$ takes a negative value in the ERTA calculations compared to the exact results from $\lambda \phi^4$ theory. It is important to note that the above observations are valid only for the choice of $\bar{\kappa}$ that is defined in the $\lambda \phi^4$ theory which has an inverse dependence on $e^\alpha$. With the ansatz for the $\tau_R(p)$ as described in Eq.~(\ref{m2.38}), $\bar{\kappa}$ is a dimensionless constant. Taking $\bar{\kappa}$ to be constant leads to $\lambda_{\pi n} = 2/(3\beta)$ which is now a positive value (also depicted in Fig.~\ref{figure1}). The negative value of $\lambda_{\pi n}$ for this particular choice of $\bar{\kappa}$ from $\lambda \phi^4$ theory arises due to its dependence on $e^\alpha$ (which modifies the terms associated with derivatives with respect to $\alpha$ in Eq.~(\ref{mm2.37})). For a one-to-one matching of ERTA results with scalar field theory predictions (where degeneracy factor is taken as 1), the first three ERTA coefficients (in the Table~\ref{t1}) differ by a factor of 1/3, even though their temperature behavior is exactly similar. Despite this, most of the second-order ERTA coefficients match exactly with those from the scalar theory. Hence, it is expected that the overall hydrodynamical evolution with the ERTA may not be significantly different from that of the scalar theory. We note that results from holography \cite{Fuini:2016qsc}, suggests that for a strongly coupled system, the high momentum modes relax faster than low momentum modes leading to a negative power of the momentum dependence in the relaxation time ($\ell<0$) and for a weakly coupled system, low momentum modes relax faster than the high momentum modes leading to ($\ell>0$). Since our results for $\ell=1$ mostly agrees with that of $\lambda \phi^4$ theory which is a weakly coupled system, they are also consistent with the observations from \cite{Fuini:2016qsc}.}

The presence of an external source current will further affect the shear viscous coefficients of the medium. In the next section, we will analyze the impact of the magnetic field on the shear viscous coefficients of the QCD medium within the ERTA.
\section{Magnetohydrodynamic shear evolution with ERTA, $\mu\neq 0$,  $B\neq 0$}\label{sec2}
In the conventional formulation of relativistic magnetohydrodynamics, the equations of fluid dynamics are coupled with Maxwell’s equations taking both the electric and magnetic fields as dynamical variables:
\begin{equation}
    \begin{aligned}
        & \partial_\mu F^{\mu\nu} = J^\nu, \\
        & \partial_\mu \Tilde{F}^{\mu\nu} = 0,
    \end{aligned}
\end{equation}
where $J^\mu$ is the electric charge four-current. The electromagnetic field tensor $F^{\mu\nu}$ can be decomposed into components parallel and perpendicular to the fluid velocity $u^\mu$ as,
\begin{equation}
    \begin{aligned}
        F^{\mu \nu}& = F^{\mu \lambda} u_\lambda u^\nu + F^{\lambda \nu} u_\lambda u^\mu + \Delta^\mu_\alpha \Delta^{\nu}_\beta F^{\alpha \beta}  \\
        & = E^\mu u^\nu - E^\nu u^\mu + \epsilon^{\mu \nu \alpha \beta}u_\alpha B_\beta.
    \end{aligned}
\end{equation}
Similarly, we can decompose its dual counter-part  $\Tilde{F}^{\mu\nu}$ as,
\begin{equation}
    \Tilde{F}^{\mu\nu}=\frac{1}{2}\epsilon^{\mu\nu\alpha\beta}F_{\alpha\beta}=B^\mu u^\nu - B^\nu u^\mu - \epsilon^{\mu\nu\alpha\beta}u_\alpha E_\beta,
\end{equation}
with $E^\mu = F^{\mu\nu}u_\nu$ and $B^\mu = F^{\mu\nu}u_\nu$. Here, $\epsilon^{\mu\nu\alpha\beta}$ is the rank-four Levi-Civita tensor with $\epsilon^{0123}=+1$ and hence $\epsilon_{\mu\nu\alpha\beta} = -\epsilon^{\mu\nu\alpha\beta}$. In the local rest frame of the fluid, $E^\mu = (0, {\bf E})$ and $B^\mu = (0, {\bf B})$ and hence these coincide with the usual electric and magnetic fields. Note that the electric and magnetic four-vectors are orthogonal to the fluid velocity $i.e$, $E^\mu u_\mu =0$ and $B^\mu u_\mu = 0$. 

{The presence of electromagnetic field introduces additional new scales into the kinetic equation, $R_L=\frac{k_{\perp}}{qB}$ which is the Larmor radius where $k_{\perp}$ is the momenta of particle perpendicular to the magnetic field. For relativistic hot plasma $k_{\perp}\sim T$ and we have $R_L=\frac{T}{qB}$ that defines the magnetic cyclotron scale. Our focus is in the regime where $r_c\ll R_L$ and $\frac{1}{T}\ll R_L$. The first inequality indicates that the impact of the magnetic field can be neglected in the collision term, and the second one implies that the Landau quantization is not required and can proceed with the classical treatment. In this regime, it is safer to assume that the magnetization pressure -$\mathcal{M}B$ (where $\mathcal{M}$ is the magnetization) is much smaller than thermodynamic pressure $P$. In other words, even though we started the analysis by considering electromagnetic fields are dynamical fields, for the case with vanishing magnetization or polarization, the fields are acting as external fields, and hence the magnetic field effects enter through the Lorentz force term in the kinetic equation~\cite{Hattori:2022hyo}. 
The microscopic description and evolution of the medium with a finite magnetization is yet not well understood from the kinetic theory point of view (the complexities lie in dealing with the dipole moment of the consistent particles of the medium). The present study focuses on the regime $\frac{1}{T}\ll R_L$ where magnetization effects are assumed to have a negligible impact.} 

In the present analysis, we work with the widely used approximation of an infinitely conducting medium, before dealing with the more complicated case of a medium with a finite electrical conductivity in a follow-up work. When the electrical conductivity is assumed to be infinite, $\sigma_E \rightarrow \infty$, the induced current $J_d^\mu = \sigma_E E^\mu \rightarrow \infty$ for any value of the electric field strength. Hence, in this limit, we demand that $E^\mu = 0$ and so in an arbitrary reference frame, ${\bf E}= - {\bf v} \times {\bf B}$. This eliminates the electric field from the equations of motion. While deriving dissipative magnetohydrodynamics from kinetic theory, dissipation is understood to be a result of collisions between the particles in the medium with the mean free path between two collisions being, $\lambda_{mfp} \sim 1/(n \sigma)$, where $\sigma$ is the scattering cross section for $2 \rightarrow 2$ collisions and $n$ is the number density. Ideal magnetohydrodynamics with an infinite electrical conductivity though assumes no collisions or an infinite $\lambda_{mfp}$. Hence, dissipative magnetohydrodynamics with infinite conductivity will not describe the complete evolution. Nevertheless, we particularly focus on the dissipation with vanishing electric field, as our current interest lies on the momentum transport (evolution of shear stress tensor) of the viscous medium.  We intend to explore the case of finite conductivity soon in near future.

Under the assumption of $E^\mu = 0$, the electromagnetic field tensor becomes,
\begin{equation}\label{3.4}
    F^{\mu\nu} \rightarrow B^{\mu\nu} = \epsilon^{\mu\nu\alpha\beta}u_\alpha B_\beta = -B b^{\mu\nu},
\end{equation}
where $B^\mu B_\mu = -B^2 $ and $b^\mu=\frac{B^\mu}{B}$ with $b^\mu u_\mu=0$ and $b^\mu b_\mu=-1$. In the presence of a magnetic field,  it is convenient to introduce a rank-two operator which projects onto a subspace orthogonal to both $u^\mu$ and $b^\mu$:
\begin{equation}
    \Xi^{\mu\nu}=\Delta^{\mu\nu}+b^\mu b^\nu.
\end{equation}
From Eq.~(\ref{3.4}), we define the antisymmetric tensor $b^{\mu\nu}=-\epsilon^{\mu\nu\alpha\beta}u_\alpha b_\beta$ which is orthogonal to both $b^\mu$ and $u^\mu$ $i.e$, $b^{\mu\nu}b_\nu=b^{\mu\nu}u_\nu=0$ with $b^{\mu\nu}b_{\mu\nu}=2$. Using the identities for $\epsilon^{\mu\nu\alpha\beta}$, we have $b^{\mu\alpha} b_{\nu\alpha} = \Xi^\mu_\nu$. Under these conventions in Maxwell’s equations, the evolution equation for the magnetic field driven by the fluid current can be described as, 
\begin{align}
    &\epsilon^{\mu\nu\alpha\beta}\left( u_\alpha \partial_\mu B_\beta + B_\beta \partial_\mu u_\alpha \right) = J^\nu,\\
    &\dot{B}^\mu + B^\mu \theta = u^\mu \partial_\nu B^\nu + B^\nu \nabla_\nu u^\mu.
\end{align}
Further, the electromagnetic energy-momentum tensor can be described as~\cite{Denicol:2018rbw},
\begin{equation}
    T^{\mu\nu}_{em} = - F^{\mu\lambda}F^{\nu}_{\;\; \: \lambda} + \frac{1}{4} g^{\mu\nu} F^{\alpha\beta}F_{\alpha\beta} =  \frac{B^2}{2} \left(u^\mu u^\nu - \Delta^{\mu\nu} - 2 b^\mu b^\nu \right).
\end{equation}
Here we have considered a non-magnetizable and non-polarizable fluid for the current study. Using Maxwell’s equations, the evolution of $T^{\mu\nu}_{em}$ can be written as,
\begin{equation}\label{3.10}
    \partial_\mu T^{\mu\nu}_{em}=-F^{\nu\lambda}J_\lambda = Bb^{\mu\lambda}J_\lambda.
\end{equation}
\subsection{Equation of motion in the presence of electromagnetic fields}
For a non-magnetizable and non-polarizable fluid, the total energy-momentum tensor can be expressed as the sum of the energy-momentum tensor of the electromagnetic field and the fluid as follows,
\begin{equation}
    T^{\mu\nu}=T^{\mu\nu}_{em}+T^{\mu\nu}_f.
\end{equation}
In the more general case, the total current is the combination of the current from the fluid ($J^\mu_f$) and an external current ($J^\mu_{ext}$) due to the motion of the spectator particles,
\begin{equation}
    J^\mu = J^\mu_f + J^\mu_{ext}.
\end{equation}
For the fluid current, $J^\mu_f = n_f N^\mu$, where $n_f$ is the local charge density in the local rest frame given by $n_f = q n$ and just like $N^\mu$, since the total charge is also conserved, we have,
\begin{equation}
    \partial_\mu J^\mu_f = 0.
\end{equation}
The external current source feeds energy and momentum into the system and hence we have,
\begin{equation}\label{3.14}
    \partial_\mu T^{\mu\nu}= - F^{\nu\lambda}J_{ext,\lambda}.
\end{equation}
The conservation equation for $T^{\mu\nu}_{em}$ as described in Eq.~(\ref{3.10}) becomes,
\begin{equation}\label{3.15}
    \partial_\mu T^{\mu\nu}_{em}=-F^{\nu\lambda}(J_{f,\lambda} + J_{ext,\lambda})
\end{equation}
Using Eq.~(\ref{3.14}) and Eq.~(\ref{3.15}), the equation of motion for the energy-momentum tensor of the fluid can be written as,
\begin{equation}\label{3.16}
    \partial_\mu T^{\mu\nu}_f = F^{\nu\lambda} J_{f,\lambda}.
\end{equation}
Note that the external current source doesn't appear in Eq.~(\ref{3.16}) which shows that the energy and momentum of the fluid is only affected by the Lorentz force exerted on the fluid currents by the electromagnetic fields. {We can decompose the fluid four-current as $J^\mu_f = n_f u^\mu + n_f^\mu$ and since in the non-resistive limit, $F^{\mu\nu}=-Bb^{\mu\nu}$ is orthogonal to the fluid velocity $u^\mu$, we can see that in the non-dissipative case, $\partial_\mu T^{\mu\nu}_{f0}=0$. Hence, the fluid's energy-momentum tensor in that case is separately conserved.} Contracting Eq.~(\ref{3.16}) with $u^\mu$ and $\Delta^{\alpha}_\nu$, we get the equations of motion at $B\neq 0$ as,
\begin{equation}
\begin{aligned}
    & u_\nu \partial_\mu T^{\mu\nu}_f = 0, \\
    & \Delta^\alpha_\nu \partial_\mu T^{\mu\nu}_f = -B b^{\alpha\lambda}n_\lambda.
\end{aligned}    
\end{equation}
These leads to the equation of motion as follows,
\begin{align}\label{3.18}
\dot{\varepsilon}+(\varepsilon+P)\theta-\pi^{\mu\nu}\sigma_{\mu\nu}&=0,\\
(\epsilon+P)\dot{u}^\mu - \nabla^\mu P + \Delta^\mu_\nu \partial_\gamma \pi^{\gamma\nu}&= -Bb^{\nu\lambda} n_\lambda,\\
\dot{n}+n\theta+\partial_{\mu}n^{\mu}&=0.\label{3m.5}
\end{align}
By substituting Eqs.~(\ref{2.6})-(\ref{2.8}) into Eqs.~(\ref{3.18})-(\ref{3m.5}), we obtain the form of $\dot{\alpha}$, $\dot{\beta}$ and $\dot{u^\mu}$ at a non-vanishing magnetic field as,
     \begin{align}\label{m3.20}
         &\dot{\alpha} = -A_n \partial_\mu n^\mu - A_\Pi \pi^{\mu\nu}\sigma_{\mu\nu}, \\
         &\dot{\beta} = \frac{\beta \theta}{3} - D_n \partial_\mu n^\mu - D_\Pi \pi^{\mu \nu} \sigma_{\mu\nu}, \\
         &\nabla^\alpha \beta = - \beta \dot{u}^\alpha + \frac{n}{\epsilon+P} \nabla^\alpha \alpha - \frac{\beta}{\epsilon+P} \Delta^\alpha_\nu \partial_\mu \pi^{\mu\nu} - \frac{\beta}{\epsilon+P}qB b^{\alpha \lambda} n_\lambda,\label{m3.22}
\end{align}
where the coefficients $A_n$, $A_\Pi$, $D_n$, and $D_\Pi$ are defined in Eq.~(\ref{m2.12}) and Eq.~(\ref{m2.13}). We will  use the above relations to derive the evolution equations for shear stress tensor $\pi^{\mu\nu}$ in the presence of a magnetic field, as discussed in the next section.
\subsection{ERTA Boltzmann equation at finite magnetic field}
The relativistic Boltzmann equation in the presence of a magnetic field can be described as,
\begin{equation}
    p^\mu \partial_\mu f - qB^{\sigma \nu} p_\nu \frac{\partial f}{\partial p^\sigma} = -\frac{(u \cdot p)}{\tau_R(x,p)} (f-f^*_0).
\end{equation}
In the strong magnetic field limit, where the magnetic field is
considered as the dominant energy scale compared to the temperature of the medium, the charged particle undergoes Landau-level kinematics. This, in turn, affects the QCD thermodynamics and the collisional kernel in the strong field regime~\cite{Hattori:2016lqx, Kurian:2019fty,Hattori:2017qih, Kurian:2017yxj}.  As we consider the case of a weak magnetic field regime where temperature is the dominant energy scale in comparison with magnetic field, we ignore the explicit dependence of the magnetic field on the thermodynamic quantities and relaxation time unlike in the case of strong field regime. However, the presence of a magnetic field introduces another expansion parameter $\chi =\frac{qB\tau_0(x)}{T}$ in addition to the Knudsen number $Kn=\tau_0\partial$. The validity of $\chi \ll 1$ is justified with the assumption that $\tau_0/R_L \ll 1$ in the magnetohydrodynamic regime~\cite{Panda:2020zhr}. With the above power counting scheme, in first order $\mathcal{O}(Kn)$, we obtain
\begin{equation}\label{m3.24}
    \delta f_{(1)} = -\frac{\tau_R}{(u \cdot p)}\left( p^\mu \partial_\mu -qBb^{\sigma \nu}p_\nu \frac{\partial}{\partial p^\sigma} \right)f_0 +  \delta f_{(1)}^*.
\end{equation}
Using Eqs.~(\ref{m3.20})-(\ref{m3.22}), the first term can be represented as,
\begin{align}\label{m3.25}
      -\frac{\tau_R}{(u \cdot p)}p^\mu \partial_\mu f_0 = \tau_R \left[\left(\frac{n}{\epsilon+P} - \frac{1}{(u \cdot p)}\right) p^\mu \nabla_\mu \alpha + \frac{\beta p^\mu p^\alpha \sigma_{\mu \alpha}}{(u \cdot p)} -\frac{\beta}{\epsilon+P}qp^\nu Bb_{\nu\sigma}n^\sigma \right] f_0.  
    \end{align}
The last term of Eq.~(\ref{m3.25}) which is magnetic field dependent, is of the order of $\mathcal{O}(\chi Kn)$,  and is not considered in first-order for evaluating the matching conditions. The second term in Eq.~(\ref{m3.24}) vanishes as $b^{\mu\nu} p_\nu \frac{\partial f_0}{\partial p^\mu} = 0$. Until $\mathcal{O}( Kn)$, the matching condition remain intact as defined in Eq.~(\ref{2.18}). With the gradient expansion method, we observed that the magnetic field does not affect the shear viscous evolution due to $\delta f^{(1)}$. This observation is consistent with that of the RTA result~\cite{Panda:2020zhr}. However, the matching condition will directly affect the second-order evolution equation of the shear stress tensor in the presence of the magnetic field. The distribution function until second-order gradient expansion  $\Delta f=\delta f_{(1)}+\delta f_{(2)}$ can be obtained from the ERTA Boltzmann equation using the Chapman-Enskog expansion as,
\begin{align}\label{3.27}
    \Delta f = &  -\frac{\tau_R}{(u \cdot p)}p^\gamma \partial_\gamma f_0-\frac{\tau_R}{(u \cdot p)}p^\gamma \partial_\gamma \delta f_{(1)}+\frac{\tau_R}{(u \cdot p)}q B b^{\sigma \nu} p_\nu \frac{\partial}{\partial p^\sigma} \delta f_{(1)} +\frac{\tau_R}{(u \cdot p)} qB b^{\sigma \nu}p_\nu \frac{\partial f_0}{\partial p^\sigma}+\Delta f_{(2)}^*, 
\end{align}
where $\delta f_{(1)}$ in the presence of a magnetic field is defined in Eq.~(\ref{m3.24}). Similar to the case of a vanishing magnetic field, we can define,
\begin{align}
     &\Delta f^*_{(2)} = \left[ -\frac{(\delta u_{(2)} \cdot p)}{T} + \frac{(u \cdot p - \mu)}{T^2} \delta T_{(2)} + \frac{\delta \mu_{(2)}}{T}\right] f_0.
\end{align}
Unlike the previous case, $\delta u_{(2)}$, $\delta T_{(2)}$, and $\delta \mu_{(2)}$ can have dependence on the magnetic field in the medium. 
\subsection{Shear stress evolution in the presence of a magnetic field}
The evolution equation of shear in the presence of a magnetic field within the ERTA framework until second-order gradient expansion can be obtained from,
\begin{align}\label{m3.28}
    \pi^{\mu\nu} = & \Delta^{\mu\nu}_{\alpha\beta}\int{dP}\,p^{\alpha}p^{\beta}\Bigg\{\underbrace{\Delta f_{(2)}^*}_{I_1} -\underbrace{\frac{\tau_R}{(u \cdot p)}p^\gamma \partial_\gamma f_0}_{I_2}-\underbrace{\frac{\tau_R}{(u \cdot p)}p^\gamma \partial_\gamma \delta f_{(1)}}_{I_3}+\underbrace{\frac{\tau_R}{(u \cdot p)}q B b^{\sigma \nu} p_\nu \frac{\partial}{\partial p^\sigma} \delta f_{(1)}}_{I_4}\nonumber \\
    &+\underbrace{\frac{\tau_R}{(u \cdot p)} qB b^{\sigma \nu}p_\nu \frac{\partial f_0}{\partial p^\sigma}}_{I_5}\Bigg\}+ f_0 \rightarrow \bar{f}_0.
\end{align}
The notation $f_0 \rightarrow \bar{f}_0$ denotes the corresponding antiparticle contributions to each term. The detailed derivation of solving each terms ($I_i\,\,\, (i=1,..,5)$) is presented in Appendix~\ref{AC}. We observed  that the direct magnetic field dependence in the shear viscous evolution is entering through the third and fourth terms ($I_3$ and $I_4$ terms). Solving Eq.~(\ref{m3.28}), we obtain  
\begin{align}\label{3.30}
    \dot{\pi}^{\langle\mu\nu\rangle} &+ \frac{\pi^{\mu\nu}}{\tau_\pi} =  2\beta_\pi \sigma^{\mu\nu} - \frac{4}{3} \pi^{\mu\nu}\theta +2\pi_\gamma^{\langle\mu}\omega^{\nu\rangle\gamma} - \tau_{\pi\pi}\pi_\gamma^{\langle\mu}\sigma^{\nu\rangle\gamma}\nonumber \\
    &-\tau_{\pi n} n^{\langle \mu}\dot{u}^{\nu\rangle}+\lambda_{\pi n} n^{\langle \mu}\nabla^{\nu\rangle}\alpha+ l_{\pi n} \nabla^{\langle\mu}n^{\nu\rangle}+\delta_{\pi B}\Delta^{\mu\nu}_{\eta \beta}qBb^{\gamma\eta}g^{\beta\rho}\pi_{\gamma\rho}\nonumber\\
    &-qB\tau_{\pi n B}\dot{u}^{\langle \mu}b^{\nu \rangle \sigma}n_\sigma-qB\lambda_{\pi n B} n_\sigma b^{\sigma \langle \mu}\nabla^{\nu \rangle}\alpha - q\tau_0\delta_{\pi n B} \nabla^{\langle \mu}\left(B^{\nu\rangle\sigma}n_\sigma\right),
\end{align}
where the coefficients $\beta_\pi$, $\tau_{\pi\pi}$, $\tau_{\pi n}$, $\lambda_{\pi n}$, and $l_{\pi n}$ are not affected due to the addition of magnetic field and is defined in Eqs.~(\ref{mm2.32})-(\ref{mm2.37}).
\begin{figure}
\begin{center}
\includegraphics[scale=0.4580]{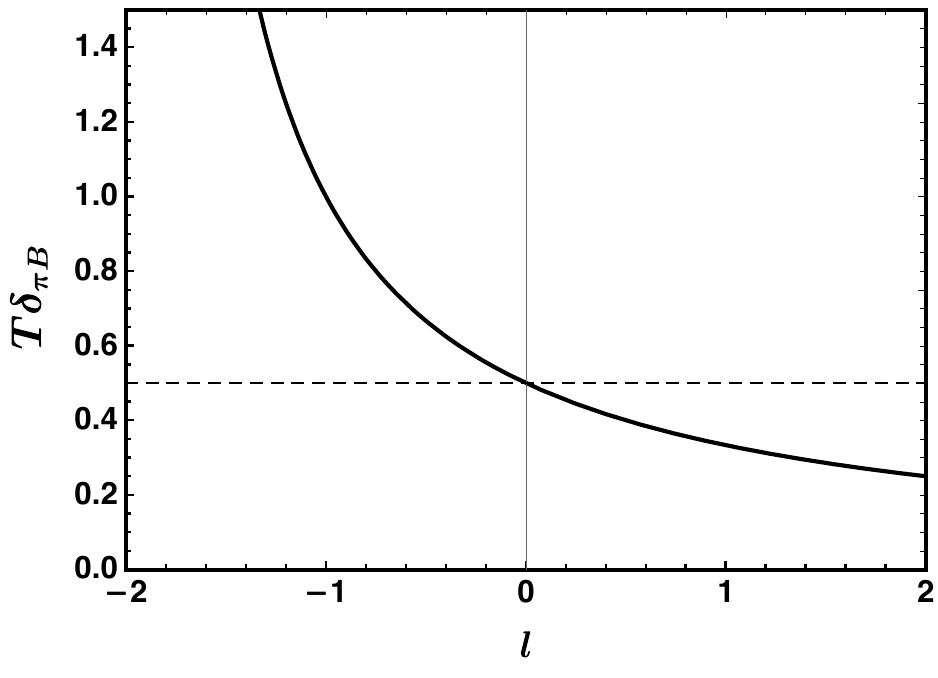}
\includegraphics[scale=0.4580]{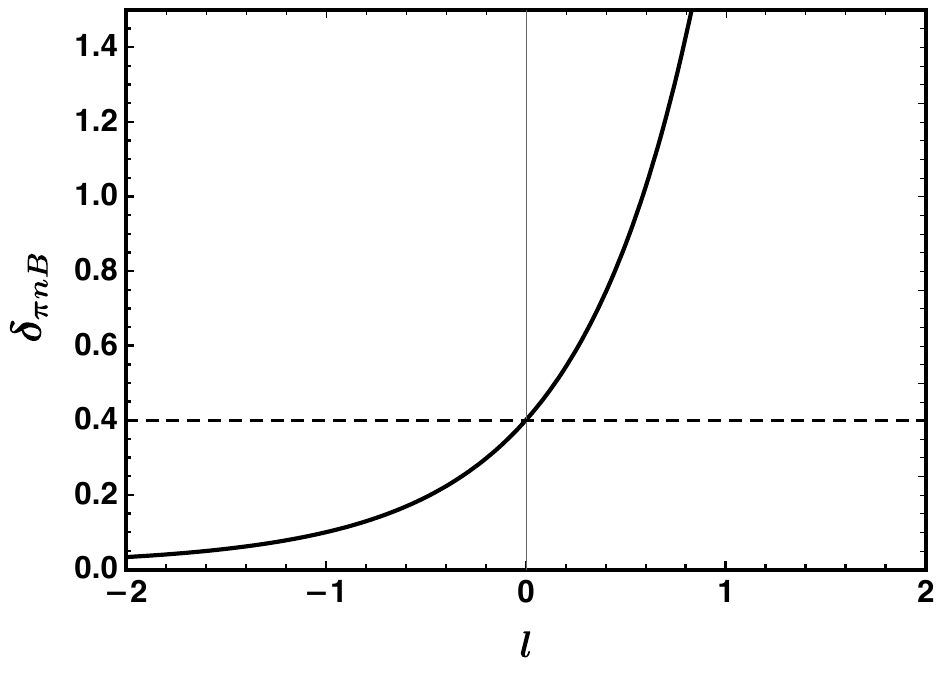}
\caption{ \small Coefficients $T\delta_{\pi B}$ (left panel) and $\delta_{\pi n B}$ (right panel) as a function of the momentum dependence parameter $\ell$ of the relaxation time. At $\ell=0$, $i.e$, in the RTA limit,  we obtain $T\delta_{\pi B}=\frac{1}{2}$ and $\delta_{\pi n B}=\frac{2}{5}$~\cite{Panda:2020zhr}.}
\label{figure2}
\end{center}
\end{figure}
The new coefficients that couple shear stress tensor with magnetic field are given by,
\begin{align}
    \delta_{\pi B}&=2\frac{\beta L_{22}^-}{\eta_0 \tau_\pi},\\
    \tau_{\pi n B}&= \frac{1}{\tau_\pi (\epsilon+P)}\Bigg[ \beta^2 N_{32}^+ -\beta M_{42}^+ + \beta(\epsilon+P)\frac{\partial \xi}{\partial \beta}L_{32}^+\Bigg],\\
    \delta_{\pi n B} &=\frac{2}{\tau_\pi \tau_0} \xi L_{32}^+,\\
    \lambda_{\pi n B}&=\frac{2}{\tau_\pi(\epsilon+P)}\Bigg[ 2\beta\tilde{\chi} L_{42}^+ +2\beta L_{32}^+  +\beta \tilde{\chi} N_{32}^++C\beta^3K_{42}^+-\beta N_{22}^- \nonumber \\
    &+(\epsilon+P)\left(\frac{\partial \xi}{\partial \alpha}+\xi\frac{\partial \xi}{\partial \beta}\right)L_{32}^++\beta\frac{\partial \xi}{\partial \beta}L_{32}^++\beta^3\frac{\partial C}{\partial \beta}K_{32}^+ - 2\beta^2K_{32}^+\Bigg],   
\end{align}
with $\xi=\frac{\beta}{\epsilon+P}$. The Eq.~(\ref{3.30}) describes the ERTA result of shear viscous evolution in the presence of a magnetic field. We compared our results with conventional RTA results and observed that our results are consistent with other studies at various limits as:
\begin{itemize}
    \item  In our estimation, we used Eq.~(\ref{m2.38}) to describe the relaxation time with $\ell$ as the parameter that controls the momentum-dependence of the $\tau_R$. At $B\neq 0$ and $\ell=0$ case, Eq.~(\ref{3.30}) reduces back to the Ref.~\cite{Panda:2020zhr}, in the massless limit. It is seen that magnetic field-dependent transport coefficients are significantly modified within the current ERTA setup. In Fig.~\ref{figure2}, we showed the impact of $\ell$ on the coefficients $\delta_{\pi B}$ and $\delta_{\pi n B}$. With an increase in the value of $\ell$, the $T\delta_{\pi B}$ decreases, however, the behaviour is quite the opposite for $\delta_{\pi n B}$. We have correctly reproduced the RTA estimations, $\delta_{\pi B}=\frac{\beta}{2}$ and $\delta_{\pi n B}=\frac{2}{5}$ respectively, at $\ell=0$~\cite{Panda:2020zhr}. The $\ell$ dependence of $\delta_{\pi B}$ will further affect the first-order anisotropic shear viscous coefficients of the magnetized medium in the Navier-Stokes limit, as shown in the next subsection. 
    \item  At $B=0$, $\mu=0$, and $\ell=0$, we note that our result is consistent with the previous study~\cite{Bhalerao:2013pza}. Also, in the limit where $B=0$, $\mu\neq 0$, and $\ell=0$, Eq.~(\ref{3.30}) reduces to that in Ref.~\cite{Jaiswal:2015mxa}. We have also reproduced the results of Ref.~\cite{Dash:2023ppc} in the case of   $B=0$, $\mu= 0$, and $\ell\neq 0$.
\end{itemize}
\subsection{The Navier-Stokes limit: Magnetic field-dependent shear coefficients}
\begin{figure}
\begin{center}
\includegraphics[scale=0.4]{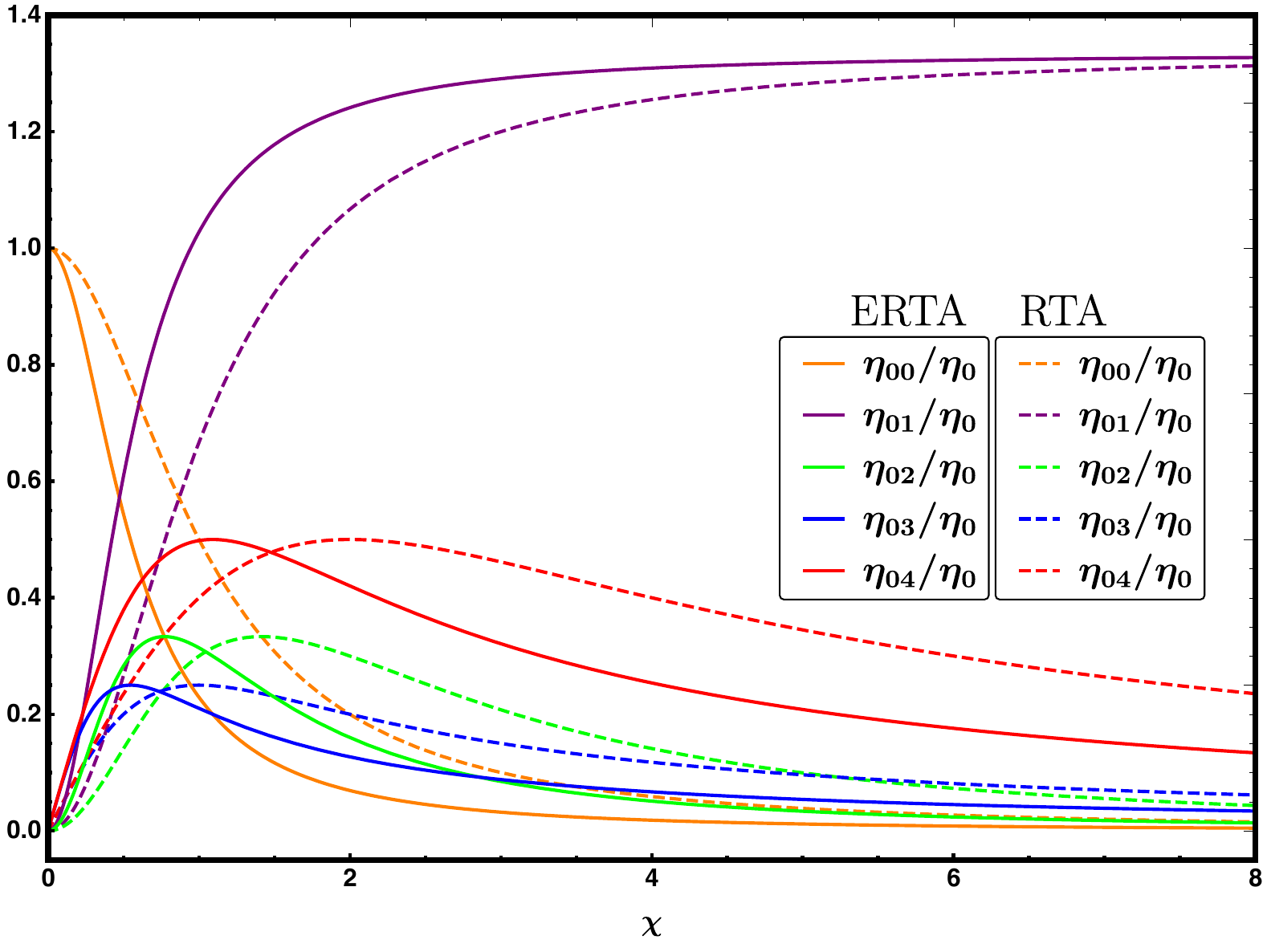}
\caption{ \small Comparison of the ERTA magnetic field-dependent shear coefficients with that of the RTA estimations. Solid lines are the ERTA calculations by choosing $\ell=0.5$ and dashed lines denote the RTA results.}
\label{figure3}
\end{center}
\end{figure}
In the formulation of magnetohydrodynamics, the magnetic field is taken to be of leading order or $\mathcal{O}(1)$ in gradients since it doesn't get screened, unlike the electric field. This is due to the absence of magnetic monopoles leading to the Bianchi identity. Keeping all the terms of  $\mathcal{O}(Kn)$  and $\mathcal{O}(\chi Kn)$ ($i.e$ first-order in gradient) in Eq.~(\ref{3.30}), we obtain the Navier-Stokes form of the constitutive relation for the shear stress tensor within ERTA as,
\begin{equation}\label{3.31}
    \left(\frac{g^{\mu\gamma}g^{\nu\rho}}{\tau_\pi}-\delta_{\pi B} \Delta^{\mu\nu}_{\eta\beta} qBb^{\gamma\eta} g^{\beta\rho}\right)\pi_{\gamma\rho}=2\beta_\pi \sigma^{\mu\nu}.
\end{equation}
{The second law of local thermodynamics along with the Curie principle leads to a general form of $\pi^{\mu\nu}$ as,}
\begin{align}\label{3.32}
    \pi^{\mu\nu}=&\Bigg[2\eta_{00} \left( \Delta^{\mu\alpha}\Delta^{\nu\beta}\right)+\eta_{01} \left(\Delta^{\mu\nu}-\frac{3}{2}\Xi^{\mu\nu} \right)\left( 
 \Delta^{\alpha\beta} - \frac{3}{2}\Xi^{\alpha\beta}\right) -2\eta_{02} \left( \Xi^{\mu\alpha}b^\nu b^\beta + \Xi^{\nu\alpha}b^\mu b^\beta\right) \nonumber\\ &-2\eta_{03} \left(\Xi^{\mu\alpha}b^{\nu\beta} + \Xi^{\nu\alpha}b^{\mu\beta}\right) +2\eta_{04}
 \left( b^{\mu\alpha}b^\nu b^\beta + b^{\nu\alpha}b^\mu b^\beta \right)\Bigg]\sigma_{\alpha\beta}.
 \end{align}
The projection operators in the presence of a magnetic field and their properties are presented in Appendix~\ref{AB}. Substituting Eq.~(\ref{3.32}) into Eq.~(\ref{3.31}), and comparing  various tensorial structures on the left-hand and the right-hand sides of Eq.~(\ref{3.31}), we obtain, 
\begin{align}
    \eta_{00} &= \frac{\beta_\pi \tau_\pi }{1+(2qB\tau_\pi\delta_{\pi B})^2},\\
    \eta_{01}&=\frac{16\beta_\pi\tau_\pi (qB\tau_\pi \delta_{\pi B})^2}{3+(2\sqrt{3}qB\tau_\pi \delta_{\pi B})^2},\\
    \eta_{02}&=\frac{3\beta_\pi\tau_\pi (qB\tau_\pi \delta_{\pi B})^2}{1+5(qB\tau_\pi \delta_{\pi B})^2+4(qB\tau_\pi\delta_{\pi B})^4},\\
    \eta_{03}&=\frac{\beta_\pi \tau_\pi (qB\tau_\pi\delta_{\pi B})}{1+(2qB\tau_\pi\delta_{\pi B})^2},\\
    \eta_{04}&=\frac{\beta_\pi \tau_\pi (qB\tau_\pi\delta_{\pi B})}{1+(qB\tau_\pi\delta_{\pi B})^2}.
\end{align}
In the limit of vanishing magnetic field $\eta_{00}\rightarrow \eta_0$, and we get back the first order constitutive relation, $\pi^{\mu\nu}=2\eta_0 \sigma^{\mu\nu}$ from Eq.~(\ref{3.32}), where $\eta_0=\beta_\pi \tau_\pi$. In Fig.~\ref{figure3}, we showed the momentum dependence of relaxation time on the anisotropic shear coefficients in the presence of a magnetic field. The relative strength of these coefficients in comparison with the case of $B=0$ is expressed in terms of $\frac{\eta_{0i}}{\eta_0}$. We observed that the RTA shear coefficients in the presence of a magnetic field are significantly modified with the ERTA setup. This arises from the ERTA-modified shear stress tensor coupling with the magnetic field, that enters through the coefficient $\delta_{\pi B}$. We also notice that at $\ell=0$, or the momentum-independent RTA case with a constant $\tau_0$, the coefficient $\delta_{\pi B}\rightarrow 2J^-_{22}/J^+_{32}$ which agrees with the value of $\delta_{\pi B}$ in Ref.~\cite{Panda:2020zhr} and $\eta_i/\eta_0$ ratios match with the one found in the RTA limit. For a finite $\ell$, the coefficients behave differently. For instance, the value of $\frac{\eta_{00}}{\eta_0}$ at $\ell=0.5$ is lower than the $\ell=0$ case, while the behavior is opposite for $\frac{\eta_{01}}{\eta_0}$. The ERTA-modified shear viscous evolution is anticipated to play a crucial role in phenomenological studies. 
\section {Conclusion and Outlook}\label{sec3}
In this study, we derived the shear viscous evolution equations and determined the associated transport coefficients for the QCD matter utilizing the recently developed ERTA (Extended Relaxation-Time Approximation) approach. The near-equilibrium distribution function is estimated by solving the extended-Boltzmann equation using a Chapman-Enskog-like gradient expansion by incorporating a momentum-dependent relaxation time. Firstly, we calculated the evolution of the shear stress tensor for a charge-conserved conformal system. Subsequently, we expanded the ERTA framework to analyze the relativistic, non-resistive, magnetohydrodynamic evolution of the shear tensor. The sensitivity of the first and second-order shear viscous coefficients to the momentum dependence of the relaxation time has been explored in this study. Notably, our finding reveals that the ERTA framework substantially alters the coupling of shear stress tensor with particle diffusion current and with magnetic field as follows:
\begin{itemize}
    \item Within the conventional RTA-based Chapman-Enskog-like gradient expansion method, it has been observed that the second-order evolution equations for the shear stress tensor and the dissipative charge current are decoupled for a charge conserved conformal system~\cite{Jaiswal:2015mxa}. In the present study within the ERTA framework, we have found new transport coefficients that introduce coupling between shear viscous evolution and the charge current. These coefficients emerge due to the momentum dependence of the thermal relaxation time at finite chemical potential. {In Ref.~\cite{Jaiswal:2016hex}, it has been noted that the inclusion of mass (going beyond the conformal limit) may also introduce a coupling with shear evolution and other dissipative terms in the medium along with the diffusion current. For the conclusive physical interpretation of this coupling, an extensive analysis of dissipative hydrodynamics for a non-conformal fluid (particles with finite mass) would be necessary.} We verified the consistency of our results with parallel studies across various limiting cases {and with the exact Boltzmann predictions.}
   
    \item In the presence of a magnetic field, we observed that the ERTA-based approach significantly modifies the shear stress tensor coupling with the magnetic field. We have compared our results with the relativistic, non-resistive magnetohydrodynamics equations within the RTA at the appropriate limits. In the Navier-Stokes limit, we obtained the magnetic field-dependent anisotropic shear coefficients. Notably, the momentum dependence of the relaxation time has a visible impact on the shear coefficients depending upon the strength of the magnetic field and the temperature of the medium.
\end{itemize}

The current study primarily focuses on the significance of momentum dependence of relaxation time on the dissipative (magneto-)hydrodynamic evolution equations. As a first step, we considered a conformal system with a parametrization of the relaxation time. The iterative way of solving ERTA Boltzmann  will become more intricate  for a resistive viscous non-conformal system. Looking forward, we intend to delve deeper into the physical implications of the coupling of shear evolution and number current by analyzing a more general system, specifically a massive, resistive, charge-conserved fluid with the ERTA framework. Notably, in the presence of a magnetic field, it has been observed that the QCD medium possesses a finite magnetization, which alters its thermodynamic and transport properties~\cite{Bali:2014kia}. Exploring the evolution equation of a medium with a non-zero magnetization presents another intriguing avenue for research. We leave these aspects for future studies.

\acknowledgments
We are thankful to  Amaresh Jaiswal, Ashutosh Dash, Samapan Bhadury, and Ankit Kumar Panda for the useful discussions. We thank Gowthama K.K. for his valuable comments on the manuscript. M.K. acknowledges the support from the Special Postdoctoral Researchers Program of RIKEN.

\appendix
\section{Definition of thermodynamic integrals and their properties}\label{AA}
In the present study, we come across the following six kinds of integrals which can be used to compute various quantities given a particular microscopic distribution function, $f$: 
\begin{equation}
    K^{\alpha_1 \alpha_2 \alpha_3 ... \alpha_n}_{(m)\pm}=\int \mathrm{dP} \frac{\tau_R(x,p)}{(u \cdot p)^m} p^{\alpha_1} p^{\alpha_2} p^{\alpha_3}....p^{\alpha_n}(f_0 \pm \bar{f}_0),
\end{equation}
\begin{equation}
    L^{\alpha_1 \alpha_2 \alpha_3 ... \alpha_n}_{(m)\pm}=\int \mathrm{dP} \frac{\tau_R^2}{(u \cdot p)^m} p^{\alpha_1} p^{\alpha_2} p^{\alpha_3}....p^{\alpha_n}(f_0 \pm \bar{f}_0),
\end{equation}
\begin{equation}
    M^{\alpha_1 \alpha_2 \alpha_3 ... \alpha_n}_{(m)\pm}=\int \mathrm{dP} \tau_R \frac{\partial \tau_R}{\partial (u \cdot p)}\frac{p^{\alpha_1} p^{\alpha_2} p^{\alpha_3}....p^{\alpha_n}}{(u \cdot p)^m} (f_0 \pm \bar{f}_0),
\end{equation}
\begin{equation}
    N^{\alpha_1 \alpha_2 \alpha_3 ... \alpha_n}_{(m)\pm}=\int \mathrm{dP} \tau_R \frac{\partial \tau_R}{\partial \beta}\frac{p^{\alpha_1} p^{\alpha_2} p^{\alpha_3}....p^{\alpha_n}}{(u \cdot p)^m} (f_0 \pm \bar{f}_0),
\end{equation}
\begin{equation}
    \bar{M}^{\; \alpha_1 \alpha_2 \alpha_3 ... \alpha_n}_{(m)\pm}=\int \mathrm{dP} \frac{\partial \tau_R}{\partial (u \cdot p)}\frac{p^{\alpha_1} p^{\alpha_2} p^{\alpha_3}....p^{\alpha_n}}{(u \cdot p)^m} (f_0 \pm \bar{f}_0),
\end{equation}
\begin{equation}
    \bar{N}^{\; \alpha_1 \alpha_2 \alpha_3 ... \alpha_n}_{(m)\pm}=\int \mathrm{dP} \frac{\partial \tau_R}{\partial (u \cdot p)}\frac{p^{\alpha_1} p^{\alpha_2} p^{\alpha_3}....p^{\alpha_n}}{(u \cdot p)^m} (f_0 \pm \bar{f}_0).
\end{equation}
Each of the above momentum moments can be decomposed in terms of the hydrodynamic degrees of freedom. For a generic moment like $Q^{\alpha_1 \alpha_2 ... \alpha_n}_{(m)\pm}$, we have the following decomposition,
\begin{align}\label{A7}
    Q^{\alpha_1 \alpha_2 ... \alpha_n}_{(m)\pm}= &Q^{(m)\pm}_{n0} u^{\alpha_1}u^{\alpha_2}u^{\alpha_3}...u^{\alpha_n} + Q^{(m)\pm}_{n1} (u^{\alpha_1}u^{\alpha_2}u^{\alpha_3}....u^{\alpha_{n-2}}\Delta^{\alpha_{n-1}\alpha_n} + \text{perm}) \nonumber\\
    & + Q^{(m)\pm}_{n2}(u^{\alpha_1}u^{\alpha_2}...u^{\alpha_{n-4}}\Delta^{\alpha_{n-3}\alpha_{n-2}}\Delta^{\alpha_{n-1}\alpha_n} + \text{perm})+..... \nonumber\\
    &....+Q^{(m)\pm}_{nq}(u^{\alpha_1}\underbrace{\Delta^{\alpha_{2}\alpha_3}\Delta^{\alpha_{4}\alpha_5}....\Delta^{\alpha_{n-1}\alpha_n}}_{q -\Delta \; \text{ terms}}+\text{perm}),
\end{align}    
where $n \ge 2q$. Each term above is made up of all the non-trivial permutations of the $\alpha$ indices according to the symmetric property of the $\Delta^{\alpha\beta}$ projector. The coefficients of these terms can then be derived using the orthogonality property of each of these terms and these are called the thermodynamic integrals which can be used to express macroscopic quantities like number density $n$, energy density $\epsilon$, pressure $P$, etc. The corresponding thermodynamic integrals are:
    \begin{align}
        &K^{(m)\pm}_{nq} = \frac{1}{(2q+1)!!}\int \mathrm{dP} \tau_R(u \cdot p)^{n-2q-m}(\Delta_{\alpha \beta} p^\alpha p^\beta)^q (f_0 \pm \bar{f}_0), \\
        &L^{(r)\pm}_{nq} = \frac{1}{(2q+1)!!}\int \mathrm{dP} \tau_R^2 (u \cdot p)^{n-2q-m}(\Delta_{\alpha \beta} p^\alpha p^\beta)^q (f_0 \pm \bar{f}_0), \\
        &M^{(m)\pm}_{nq} = \frac{1}{(2q+1)!!}\int \mathrm{dP} \tau_R \frac{\partial \tau_R}{\partial (u\cdot p)} (u \cdot p)^{n-2q-m}(\Delta_{\alpha \beta} p^\alpha p^\beta)^q (f_0 \pm \bar{f}_0), \label{A10}\\
        &N^{(m)\pm}_{nq} = \frac{1}{(2q+1)!!}\int \mathrm{dP} \tau_R \frac{\partial \tau_R}{\partial \beta} (u \cdot p)^{n-2q-m}(\Delta_{\alpha \beta} p^\alpha p^\beta)^q (f_0 \pm \bar{f}_0), \\
        &\bar{M}^{(m)\pm}_{nq} = \frac{1}{(2q+1)!!}\int \mathrm{dP} \frac{\partial \tau_R}{\partial (u\cdot p)} (u \cdot p)^{n-2q-m}(\Delta_{\alpha \beta} p^\alpha p^\beta)^q (f_0 \pm \bar{f}_0),  \\
        &\bar{N}^{(m)\pm}_{nq} = \frac{1}{(2q+1)!!}\int \mathrm{dP} \frac{\partial \tau_R}{\partial \beta} (u \cdot p)^{n-2q-m}(\Delta_{\alpha \beta} p^\alpha p^\beta)^q (f_0 \pm \bar{f}_0).       
    \end{align}
In the massless case, we have the following relation for each of these thermodynamic integrals,
\begin{equation}
    Q^{(m)\pm}_{n,q}=-\left(\frac{1}{2q+1}\right)Q^{(m)\pm}_{n,q-1}.
\end{equation}
The $M^{(m)\pm}_{nq}$ and $N^{(m)\pm}_{nq}$ integrals can be expressed in $L^{(m)\pm}_{nq}$ integrals as follows,
\begin{align}
    M^{(m)\pm}_{nq} &= \frac{1}{2T} L^{(m)\pm}_{nq} - \frac{n+1}{q} L^{(m)\pm}_{n-1,q}, \; \; n > -1\\
    2 N_{n,q}^{(m)\pm} &= L^{(m)\pm}_{n+1,q} - nTL^{(m)\pm}_{n,q}.
\end{align}
We also have the following relations for each of the thermodynamic integrals:
\begin{align}
    Q^{(m)\pm}_{n,q} &= T \left[ -Q^{(m)\pm}_{n-1,q-1} + (n-2q)Q^{(m)\pm}_{n-1,q} \right],\\
    \bar{N}^{(m)\pm}_{nq}&= K^{(m)\pm}_{n+1,q}-(n+1)TK^{(m)\pm}_{nq}.
\end{align}
The above relations can be used to check the limiting cases when we take $\tau_R \rightarrow \tau_0$ as well as for $B \rightarrow 0$.
\section{Projection operators in the presence of a magnetic field}\label{AB}
{To ensure semi-positive entropy production, we can represent the constitutive relation as,} 
\begin{equation}
    \pi^{\mu\nu}=\eta^{\mu\nu\alpha\beta}\sigma_{\alpha\beta},
\end{equation}
where $\eta^{\mu\nu\alpha\beta}$ follows the properties of $\pi^{\mu\nu}$, $i.e$, $\eta^{\mu\nu\alpha\beta}$ is  traceless, orthogonal to $u^\mu$, and is symmetric under the exchange of $\mu$ and $\nu$ indices. Also, $\eta^{\mu\nu\alpha\beta}$ is constructed from the tensors $u^\mu$, $b^\mu$, $g^{\mu\nu}$ and $b^{\mu\nu}$. This can be achieved by using the following tensorial structures~\cite{Huang:2011dc}:
\begin{align}
        & (i) \quad \Delta^{\mu\nu}=g^{\mu\nu}-u^\mu u^\nu,\nonumber \\
        & (ii) \quad \Xi^{\mu\nu}=\Delta^{\mu\nu} + b^\mu b^\nu, \nonumber\\
        & (iii) \quad b^\mu b^\nu, \nonumber\\
        & (iv) \quad  b^{\mu\nu},
\end{align}
where $\Delta^{\mu\nu}$ projects onto a subspace orthogonal to $u^\mu$, and $\Xi^{\mu\nu}$ projects onto the two dimensional subspace orthogonal to both $u^\mu$ and $b^\mu$. Using these we can construct a set of five tensorial combinations that follows the properties of $\pi^{\mu\nu}$ as,
\begin{align}
        & (i) \quad \Delta^{\mu \alpha} \Delta^{\nu \beta}+\Delta^{\mu \beta} \Delta^{\nu \alpha}-\frac{2}{3} \Delta^{\mu \nu} \Delta^{\alpha \beta},\nonumber \\
        & (ii) \quad \left(\Delta^{\mu \nu}-\frac{3}{2} \Xi^{\mu \nu}\right)\left(\Delta^{\alpha \beta}-\frac{3}{2} \Xi^{\alpha \beta}\right),\nonumber \\
        & (iii) \quad -\Xi^{\mu \alpha} b^\nu b^\beta-\Xi^{\nu \beta} b^\mu b^\alpha-\Xi^{\mu \beta} b^\nu b^\alpha-\Xi^{\nu \alpha} b^\mu b^\beta,\nonumber\\
        & (iv) \quad -\Xi^{\mu \alpha} b^{\nu \beta}-\Xi^{\nu \beta} b^{\mu \alpha}-\Xi^{\mu \beta} b^{\nu \alpha}-\Xi^{\nu \alpha} b^{\mu \beta},\nonumber \\
        & (iv) \quad b^{\mu \alpha} b^\nu b^\beta+b^{\nu \beta} b^\mu b^\alpha+b^{\mu \beta} b^\nu b^\alpha+b^{\nu \alpha} b^\mu b^\beta.
\end{align}
The coefficients of these combinations are the various $\eta_{0i}$ components of the shear coefficient. Using these along with the symmetric nature of $\sigma_{\alpha\beta}$, we obtain,
\begin{align}\label{B.4}
    \pi^{\mu\nu}=&\Bigg(2\eta_{00} \left( \Delta^{\mu\alpha}\Delta^{\nu\beta}\right)+\eta_{01} \left(\Delta^{\mu\nu}-\frac{3}{2}\Xi^{\mu\nu} \right)\left( 
 \Delta^{\alpha\beta} - \frac{3}{2}\Xi^{\alpha\beta}\right) -2\eta_{02} \left( \Xi^{\mu\alpha}b^\nu b^\beta + \Xi^{\nu\alpha}b^\mu b^\beta\right) \nonumber\\ &-2\eta_{03} \left(\Xi^{\mu\alpha}b^{\nu\beta} + \Xi^{\nu\alpha}b^{\mu\beta}\right)+2\eta_{04} \left( b^{\mu\alpha}b^\nu b^\beta + b^{\nu\alpha}b^\mu b^\beta \right)\Bigg)\sigma_{\alpha\beta}.
 \end{align}
\section{Derivation of second-order magnetohydro evolution of shear stress}\label{AC}
From Eq.~(\ref{m3.28}), we define the shear stress tensor as the sum of these five integrals as,
\begin{equation}
    \pi^{\mu\nu}_{(2)}=\mathcal{I}_1+\mathcal{I}_2+\mathcal{I}_3+\mathcal{I}_4+\mathcal{I}_5.
\end{equation}
Where the $I_i$ terms can be described as,
\begin{align}
    \mathcal{I}_1 &= \Delta^{\mu\nu}_{\alpha\beta} \int \mathrm{dP} \left[ -\frac{(\Delta u_{(2)} \cdot p)}{T} + \frac{(u \cdot p - \mu)}{T^2} \Delta T_{(2)} + \frac{\Delta \mu_{(2)}}{T}\right] f_0 \hspace{0.3cm} + f_0 \rightarrow \bar{f}_0 \\
    \mathcal{I}_2 &= -\Delta^{\mu\nu}_{\alpha\beta} \int \mathrm{dP} p^\alpha p^\beta \frac{\tau_R}{(u \cdot p)} p^\gamma \partial_\gamma f_0 \hspace{0.3cm} + f_0 \rightarrow \bar{f}_0\\
    \mathcal{I}_3&=-\Delta^{\mu\nu}_{\alpha\beta} \int \mathrm{dP} p^\alpha p^\beta \frac{\tau_R}{(u \cdot p)}p^\gamma \partial_\gamma  \Bigg\{-\frac{C p^l \nabla_l \alpha}{T^2} f_0 + \tau_R \Bigg[\left(\frac{n}{\epsilon+P} - \frac{1}{(u \cdot p)}\right) p^l \nabla_l \alpha \nonumber\\
    &+ \frac{\beta p^l p^m \sigma_{l m}}{(u \cdot p)} - \frac{\beta}{\epsilon+P} qp^k Bb_{k\sigma}n^\sigma \Bigg] f_0\Bigg\}\hspace{0.3cm} + f_0 \rightarrow \bar{f}_0 \\
    \mathcal{I}_4 &=\Delta^{\mu\nu}_{\alpha\beta} \int \mathrm{dP} p^\alpha p^\beta \frac{\tau_R}{(u \cdot p)}qBb^{\sigma k}p_k \frac{\partial}{\partial p^\sigma}\Bigg\{-\frac{C p^l \nabla_l \alpha}{T^2} f_0 + \tau_R \Bigg[\left(\frac{n}{\epsilon+P} - \frac{1}{(u \cdot p)}\right) p^l \nabla_l \alpha \nonumber\\
    &+ \frac{\beta p^l p^m \sigma_{l m}}{(u \cdot p)} - \frac{\beta}{\epsilon+P} qp^\nu Bb_{\sigma\nu}n^\sigma \Bigg] f_0\Bigg\}\hspace{0.3cm} + f_0 \rightarrow \bar{f}_0\\
    \mathcal{I}_5 &=\Delta^{\mu\nu}_{\alpha\beta} \int \mathrm{dP} p^\alpha p^\beta \frac{\tau_R}{(u \cdot p)} qB b^{k\sigma}p_k \frac{\partial f_0}{\partial p^\sigma}\hspace{0.3cm} + f_0 \rightarrow \bar{f}_0
\end{align}
For analytical simplicity, we evaluate the integrals only for particles, but the final expressions are presented as the combined result for particles and antiparticles.  Solving $\mathcal{I}_1$, we have
\begin{equation}\label{C7}
    \mathcal{I}_1 = \Delta^{\mu\nu}_{\alpha\beta} \int \mathrm{dP} p^\alpha p^\beta \left[ -\frac{(\Delta u_{(2)} \cdot p)}{T} + \frac{(u \cdot p - \mu)}{T^2} \Delta T_{(2)} + \frac{\Delta \mu_{(2)}}{T}\right] f_0, 
\end{equation}
 where $\Delta u_{(2)}$, $\Delta T_{(2)}$ and $\Delta \mu_{(2)}$ contains terms till second-order in gradients. Using the properties of thermodynamic integrals and four-rank symmetric tensor, Eq.~(\ref{C7}) further leads to,
\begin{align}
    \mathcal{I}_1=&-\Delta^{\mu\nu}_{\alpha\beta} I^{\alpha\beta\gamma}_{+}\frac{\Delta u_{\gamma(2)}}{T} + \Delta^{\mu\nu}_{\alpha\beta}I^{\alpha\beta\gamma}_{+}u_\gamma \frac{\Delta T_{(2)}}{T^2} - \Delta^{\mu\nu}_{\alpha\beta}I^{\alpha\beta}_{-} \mu\frac{\Delta T_{(2)}}{T^2} + \Delta^{\mu\nu}_{\alpha\beta}I^{\alpha\beta}_{-}\frac{\Delta \mu_{(2)}}{T}\nonumber\\
    &=0.
\end{align}
This indicates that the second-order matching condition doesn't affect the shear calculations. However, as we will see, the first-order matching condition will contribute to the shear viscous evolution. Now, let us evaluate $\mathcal{I}_2$,
\begin{align}
\mathcal{I}_2 &= -\Delta^{\mu\nu}_{\alpha\beta} \int \mathrm{dP} p^\alpha p^\beta \frac{\tau_R}{(u \cdot p)} p^\gamma \partial_\gamma f_0 \nonumber\\
&= -\Delta^{\mu\nu}_{\alpha\beta}\int \mathrm{dP} p^\alpha p^\beta \frac{\tau_R}{(u \cdot p)}p^\gamma \Bigg\{-\partial_\gamma\beta (u \cdot p) - \beta p^k \partial_\gamma u_k + \partial_\gamma \alpha \Bigg\}f_0 \nonumber\\
&=\Delta^{\mu\nu}_{\alpha \beta}K^{\alpha\beta\gamma}_+ \partial_\gamma \beta+\Delta^{\mu\nu}_{\alpha\beta}K^{\alpha\beta\gamma k}_{(1)+}\beta \partial_\gamma u_k+\Delta^{\mu\nu}_{\alpha \beta}K^{\alpha\beta\gamma}_{(1)-} \partial_\gamma \alpha \nonumber\\
&= 2K^{(1)+}_{42}\beta \sigma^{\mu\nu}=\pi^{\mu\nu}_{(1)}.
\end{align}
We observe that $\mathcal{I}_2$ is not affected by either the magnetic field or any second order term coming from $\nabla^\mu \beta$. Now, we evaluate $\mathcal{I}_3$:

\begin{align}
    \mathcal{I}_3&=-\Delta^{\mu\nu}_{\alpha\beta} \int \mathrm{dP} p^\alpha p^\beta \frac{\tau_R}{(u \cdot p)}p^\gamma \partial_\gamma  \Bigg\{-\frac{C p^l \nabla_l \alpha}{T^2} f_0 + \tau_R \Bigg[\left(\frac{n}{\epsilon+P} - \frac{1}{(u \cdot p)}\right) p^l \nabla_l \alpha \nonumber\\
    &+ \frac{\beta p^l p^m \sigma_{l m}}{(u \cdot p)} - \frac{\beta}{\epsilon+P} qp^k Bb_{k\sigma}n^\sigma \Bigg] f_0\Bigg\},\nonumber\\
    &=\mathcal{I}_{3A}+\mathcal{I}_{3B}+\mathcal{I}_{3C}, 
\end{align}
where,
\begin{align}
\mathcal{I}_{3A}&=\Delta^{\mu\nu}_{\alpha\beta} \int \mathrm{dP} p^\alpha p^\beta \tau_R D  \Bigg\{\frac{C p^l \nabla_l \alpha}{T^2} f_0 -\tau_R \Bigg[\left(\frac{n}{\epsilon+P} - \frac{1}{(u \cdot p)}\right) p^l \nabla_l \alpha +\frac{\beta p^l p^m \sigma_{l m}}{(u \cdot p)}\Bigg] f_0 \Bigg\},\nonumber\\
\mathcal{I}_{3B}&=\Delta^{\mu\nu}_{\alpha\beta} \int \mathrm{dP} p^\alpha p^\beta \frac{\tau_R}{(u \cdot p)} p^\gamma \nabla_\gamma  \Bigg\{\frac{C p^l \nabla_l \alpha}{T^2} f_0 -\tau_R \Bigg[\left(\frac{n}{\epsilon+P} - \frac{1}{(u \cdot p)}\right) p^l \nabla_l \alpha +\frac{\beta p^l p^m \sigma_{l m}}{(u \cdot p)}\Bigg] f_0 \Bigg\},\nonumber\\
\mathcal{I}_{3C}&=\Delta^{\mu\nu}_{\alpha\beta}\int \mathrm{dP} p^\alpha p^\beta \frac{\tau_R}{(u \cdot p)}p^\gamma \partial_
\gamma\left\{\frac{\beta}{\epsilon+P}\tau_R qp^k B b_{k\sigma}n^\sigma f_0\right\}.\nonumber
\end{align}
We obtain $\mathcal{I}_{3A}$ and $\mathcal{I}_{3B}$ as,
\begin{align}
    \mathcal{I}_{3A}&=-\Delta^{\mu\nu}_{\alpha\beta}D[2\beta \sigma^{\alpha \beta} L_{32}^+]+\dot{u}^{\langle \mu}\nabla^{\nu \rangle}\alpha \Bigg[ 2C\beta^2 K^+_{31} - \frac{2n}{\epsilon+P} L_{31}^+ + 2 L_{21}^- - 2C\beta^2\bar{M}^+_{42} \nonumber\\
    &+ \frac{2n}{\epsilon+P}M^+_{42} - 2 M^-_{32}\Bigg] + \frac{2}{3}\beta^2 \theta \sigma^{\langle \mu \nu \rangle} N_{32}^+, \nonumber\\
    \mathcal{I}_{3B}&=2\nabla^{\langle \mu}\left( C\beta^2K_{32}^+ - \frac{n}{\epsilon+P}L^+_{32}+L_{22}^-\right)\nabla^{\nu}\alpha -2\nabla^{\langle \mu}\alpha\left[\frac{\beta}{\epsilon+P} q B b^{\nu \rangle \sigma}n_\sigma \right]\Bigg(C\beta^2 \bar{N}^+_{32} \nonumber\\
    &- \frac{n}{\epsilon+P}N^+_{32}+\frac{N_{22}^-}{2}\Bigg)-\left(8\beta \sigma^{\langle \mu \gamma}\sigma^{\nu\rangle}_\gamma +\frac{14}{3}\theta \beta \sigma^{\mu\nu}\right)\left(M_{63}^{(2)+}+L_{63}^{(3)+}\right)+4\beta\sigma^{\langle\mu}_\gamma \omega^{\nu\rangle \gamma}L_{32}^+ \nonumber \\
    &-4\sigma^{\langle \mu \gamma}\sigma^{\nu \rangle}_\gamma \beta L_{32}^+-\frac{10}{3}\theta \beta \sigma^{\mu\nu}L_{32}^+ +\nabla^{\langle  \mu}\alpha \nabla^{\nu\rangle } \alpha \Bigg( \frac{2n^2}{(\epsilon+P)^2}N_{32}^+ 
    -\frac{2C\beta^2 n}{(\epsilon+P)} \bar{N}_{32}^+ -\frac{n}{\epsilon+P}N_{22}^-\Bigg) \nonumber \\
    &+\dot{u}^{\langle \mu} \nabla^{\nu \rangle}\alpha  \left(2C\beta^3 \bar{N}^+_{32}-\frac{2n\beta}{\epsilon+P}N^+_{32}+\beta N_{22}^-\right).
\end{align}
Further, we express $\mathcal{I}_{3C}$ as follows,
\begin{align}
\mathcal{I}_{3C}&=\Delta^{\mu\nu}_{\alpha\beta}\int \mathrm{dP} p^\alpha p^\beta \frac{\tau_R}{(u \cdot p)}p^\gamma (\partial_
\gamma \tau_R)\left\{\frac{\beta}{\epsilon+P}\tau_R qp^k B b_{k\sigma}n^\sigma f_0\right\}+\Delta^{\mu\nu}_{\alpha\beta}\int \mathrm{dP} p^\alpha p^\beta \frac{\tau_R^2}{(u \cdot p)}p^k p^\gamma \partial_
\gamma\left\{\frac{\beta}{\epsilon+P} q B b_{k\sigma}n^\sigma f_0\right\}.\nonumber
\end{align}
Solving this, we obtain
\begin{align}
    \mathcal{I}_{3C}&=2 \dot{u}^{\langle \mu}\left[\frac{\beta}{\epsilon+P} q B b^{\nu \rangle \sigma}n_\sigma \right]\left(M_{42}^++L_{31}^+ - 2M_{31}^+ + \beta N_{32}^+\right)+2\nabla^{\langle \mu}\left(L_{32}^+\frac{\beta}{\epsilon+P} q B b^{\nu \rangle \sigma}n_\sigma\right) \nonumber \\
    & + 2 N_{32}^+ \Big[ \frac{\beta}{\epsilon+P}qB b^{\langle  \mu\sigma}n_\sigma \Big(\frac{\beta}{\epsilon+P}qBb^{\nu\rangle \gamma}n_\gamma\Big) \Big]- \frac{2n}{\epsilon+P} \Big[ \frac{\beta}{\epsilon+P}qB b^{\langle  \mu\sigma}n_\sigma \Big] \nabla^{\nu \rangle }\alpha N_{32}^+.\nonumber\\
\end{align}
Hence, we obtain the expression for $\mathcal{I}_3$ as,
\begin{align}
    \mathcal{I}_3&=-\Delta^{\mu\nu}_{\alpha\beta}D[2\beta \sigma^{\alpha \beta} L_{32}^+]+\dot{u}^{\langle \mu}\nabla^{\nu \rangle}\alpha \Bigg[ 2C\beta^2 K^+_{31} - \frac{2n}{\epsilon+P} L_{31}^+ + 2 L_{21}^- - 2C\beta^2\bar{M}^+_{42} \nonumber\\
    &+ \frac{2n}{\epsilon+P}M^+_{42} - 2 M^-_{32}\Bigg] + \frac{2}{3}\beta^2 \theta \sigma^{\langle \mu \nu \rangle} N_{32}^++2\nabla^{\langle \mu}\left( C\beta^2K_{32}^+ - \frac{n}{\epsilon+P}L^+_{32}+L_{22}^-\right)\nabla^{\nu}\alpha \nonumber \\
    &-2\nabla^{\langle \mu}\left[\frac{\beta}{\epsilon+P} q B b^{\nu \rangle \sigma}n_\sigma \right]\left(C\beta^2 \bar{N}^+_{32} - \frac{n}{\epsilon+P}N^+_{32}+\frac{N_{22}^-}{2}\right)-\left(8\beta \sigma^{\langle \mu \gamma}\sigma^{\nu\rangle}_\gamma +\frac{14}{3}\theta \beta \sigma^{\mu\nu}\right)\Big(M_{63}^{(2)+} \nonumber\\
    &+L_{63}^{(3)+}\Big)+\nabla^{\langle  \mu}\alpha \nabla^{\nu\rangle } \alpha \Bigg( \frac{2n^2}{(\epsilon+P)^2}N_{32}^+ 
    -\frac{2C\beta^2 n}{(\epsilon+P)} \bar{N}_{32}^+ -\frac{n}{\epsilon+P}N_{22}^-\Bigg) + \dot{u}^{\langle \mu} \nabla^{\nu \rangle}\alpha\Big(2C\beta^3 \bar{N}^+_{32} \nonumber \\
    &-\frac{2n\beta}{\epsilon+P}N^+_{32}+\beta N_{22}^-\Big)+4\beta\sigma^{\langle\mu}_\gamma \omega^{\nu\rangle \gamma}L_{32}^+-4\sigma^{\langle \mu \gamma}\sigma^{\nu \rangle}_\gamma \beta L_{32}^+-\frac{10}{3}\theta \beta \sigma^{\mu\nu}L_{32}^+ \nonumber\\
    &+2 \dot{u}^{\langle \mu}\left[\frac{\beta}{\epsilon+P} q B b^{\nu \rangle \sigma}n_\sigma\right]\Bigg(M_{42}^++L_{31}^+ - 2M_{31}^+ + \beta N_{32}^+\Bigg)+2\nabla^{\langle \mu}\left(L_{32}^+\frac{\beta}{\epsilon+P} q B b^{\nu \rangle \sigma}n_\sigma\right) \nonumber \\
    & +2 N_{32}^+ \Big[ \frac{\beta}{\epsilon+P}qB b^{\langle  \mu\sigma}n_\sigma \Big(\frac{\beta}{\epsilon+P}qBb^{\nu\rangle \gamma}n_\gamma\Big) \Big] -\frac{2n}{\epsilon+P} \Big[ \frac{\beta}{\epsilon+P}qB b^{\langle  \mu\sigma}n_\sigma \Big] \nabla^{\nu \rangle }\alpha N_{32}^+.
\end{align}
Next, we evaluate $\mathcal{I}_4$ which have direct magnetic field contributions,
\begin{align}
    \mathcal{I}_4 &=\Delta^{\mu\nu}_{\alpha\beta} \int \mathrm{dP} p^\alpha p^\beta \frac{\tau_R}{(u \cdot p)}qBb^{\sigma k}p_k \frac{\partial}{\partial p^\sigma}\Bigg\{-\frac{C p^l \nabla_l \alpha}{T^2} f_0 + \tau_R \Bigg[\left(\frac{n}{\epsilon+P} - \frac{1}{(u \cdot p)}\right) p^l \nabla_l \alpha \nonumber\\
    &+ \frac{\beta p^l p^m \sigma_{l m}}{(u \cdot p)} - \frac{\beta}{\epsilon+P} qp^\nu Bb_{\sigma\nu}n^\sigma \Bigg] f_0\Bigg\}.
\end{align}
Using the properties $\frac{\partial f_0}{\partial p^\sigma} \propto u_\sigma$, $\frac{\partial \tau_R}{\partial p^\sigma} \propto u_\sigma$ and $\frac{\partial p^l}{\partial p^\sigma} = \delta^l_\sigma$, we obtain,
\begin{align}
    \mathcal{I}_4&=\Delta^{\mu\nu}_{\alpha\beta}\int \mathrm{dP} p^\alpha p^\beta p^k\frac{\tau_R}{(u \cdot p)} qB b^\sigma_{\; \; k} \left\{-\frac{C\nabla_\sigma \alpha}{T^2}+\frac{u_\sigma}{T^2}-\frac{\tau_R}{(u \cdot p)} \partial_\sigma +\frac{\tau_R \beta}{(u \cdot p)}p^l (\partial_l u_\sigma)\right\}f_0 \nonumber\\
    &=\left(4\beta qB b^{\sigma\langle\mu}\sigma^{\nu\rangle}_\sigma + \frac{4}{3}\theta \beta qB b^{\langle \mu\nu \rangle}\right)L_{22}^-. 
\end{align}
Here, we have employed the relation, $\frac{\partial}{\partial p^\sigma}(\partial_l f_0) = -\beta(\partial_l u_\sigma)f_0$.  We can also show that $b^{\langle \mu \nu \rangle} =0$. Hence, $\mathcal{I}_4$ can be written as,
\begin{align}
    \mathcal{I}_4 = 4\beta qB b^{\sigma\langle\mu}\sigma^{\nu\rangle}_\sigma L_{22}^-.
\end{align}
The term $\mathcal{I}_5$ vanishes since $\frac{\partial f_0}{\partial p^\sigma} \propto u_\sigma$ and $b^{k\sigma}u_\sigma =0$. Adding all the parts above we get the following expression for the second-order evolution equation for shear tensor in the presence of a magnetic field within the ERTA as:
\begin{equation}
    \begin{aligned}
    \pi^{\mu\nu}_{(2)}&=\mathcal{I}_1+\mathcal{I}_2+\mathcal{I}_3+\mathcal{I}_4+\mathcal{I}_5 \nonumber\\
        &= 2 \eta \sigma^{\mu\nu} - \Delta^{\mu\nu}_{\alpha \beta} D [ 2 \beta \sigma^{\alpha \beta} L^+_{32}] + \dot{u}^{\langle  \mu}\nabla^{\nu\rangle } \alpha \Bigg[  2C\beta^2 K^+_{31} - \frac{2n}{\epsilon+P} L_{31}^+ + 2 L_{21}^- \\
        & - 2C\beta^2\bar{M}^+_{42} + \frac{2n}{\epsilon+P}M^+_{42} - 2 M^-_{32} - \frac{2n\beta}{\epsilon+P} N_{32}^+ + 2C\beta^3 \bar{N}^+_{32} +\beta N^-_{22}\Bigg] \\
        & + \nabla^{\langle  \mu}\Bigg( 2C\beta^2 K^+_{32} - \frac{2n}{\epsilon+P} L_{32}^+ + 2 L_{22}^- \Bigg)\nabla^{\nu\rangle }\alpha + \nabla^{\langle  \mu}\alpha \nabla^{\nu\rangle } \alpha \Bigg( \frac{2n^2}{(\epsilon+P)^2}N_{32}^+ \\
        & -\frac{2C\beta^2 n}{(\epsilon+P)} \bar{N}_{32}^+ -\frac{n}{\epsilon+P}N_{22}^-\Bigg) + \theta \sigma^{\mu\nu} \Bigg( \frac{2\beta^2}{3}N_{32}^+ - \frac{4\beta}{3} L_{32}^+ - 2\beta L_{32}^+ - \frac{14 \beta}{3} M_{43}^+ - \frac{14 \beta}{3}L_{33}^+\Bigg) \\
        & + 4\beta L_{32}^+ \sigma^{\langle  \mu}_\gamma \omega^{\nu\rangle \gamma} - \sigma^{\langle  \mu \gamma}\sigma^{\nu\rangle }_\gamma \Bigg( 4\beta L_{32}^+ + 8\beta M_{43}^+ + 8\beta L_{33}^+ \Bigg) - qB \nabla^{\langle  \mu}\alpha b^{\nu\rangle }_{\; \; \sigma}n^\sigma \Bigg( \frac{2n\beta}{(\epsilon+P)^2}N_{32}^+\\
        &-\frac{2C\beta^3}{(\epsilon+P)} \bar{N}_{32}^+ -\frac{\beta}{\epsilon+P}N_{22}^- \Bigg)+ 2\dot{u}^{\langle  \mu} \Bigg[ \frac{\beta}{\epsilon+P} qBb^{\nu\rangle \sigma}n_\sigma \Bigg]\Bigg( M_{42}^+ + L_{31}^+ - 2M_{31}^+ +\beta N_{32}^+ \Bigg) \\
        & + 2 N_{32}^+ \Big[ \frac{\beta}{\epsilon+P}qB b^{\langle  \mu\sigma}n_\sigma \Big(\frac{\beta}{\epsilon+P}qBb^{\nu\rangle \gamma}n_\gamma\Big) \Big] - \frac{2n}{\epsilon+P} \Big[ \frac{\beta}{\epsilon+P}qB b^{\langle  \mu\sigma}n_\sigma \Big] \nabla^{\nu \rangle }\alpha N_{32}^+\\
        & +2 \nabla^{\langle  \mu}\Bigg( L_{32}^+ \frac{\beta}{\epsilon+P}qBb^{\nu\rangle \sigma}n_\sigma \Bigg) - 4\beta q B \sigma^{\langle  \mu \gamma} b^{\nu \rangle }_{\; \; \gamma} L_{22}^-.
    \end{aligned}
\end{equation}
In the limit of $\ell=0$, $i.e$, the conventional RTA framework with a finite magnetic field, the evolution equation of shear stress tensor reduces to the result in Ref.~\cite{Panda:2020zhr}, in the massless limit. We also observe that for the case with $B=0$, $\mu=0$, and $\ell \neq 0$, the above expression reduces to the form of second-order shear stress evolution as given in the Ref.~\cite{Dash:2023ppc}. 

{\section{Realization of conservation laws in the second-order ERTA framework}\label{AD}
Here, we show that the energy-momentum and number conservation laws are indeed satisfied in the ERTA framework up to the second order. For simplicity, we only consider a system of particles but the generalization to anti-particles is straightforward. The Boltzmann equation in the ERTA model is,
\begin{equation}
    p^\mu \partial_\mu f = -\frac{(u \cdot p)}{\tau_R} (\delta f - \delta f^*)
\end{equation}
\subsubsection{Number Conservation: $\partial_\mu N^\mu=0$}
We start with the $N^\mu$ conservation. Zeroth momentum moment of the Boltzmann equation gives rise to the number conservation law. To this end, we consider the zeroth moment of
the collision kernel and can be defined as follows,
\begin{equation}
    \int \mathrm{dP} C[f] = -\int \mathrm{dP} \frac{(u \cdot p)}{\tau_R} (\delta f - \delta f^*).
\end{equation}
The above equation up to second order in gradients can be evaluated using the expression for $\Delta f_{(2)} - \Delta f_{(2)}^*$ from Eq.~(\ref{2.24}), 
\begin{equation}
\begin{aligned}
    \int \mathrm{dP} C[f] &= \int \mathrm{dP} p^\alpha \partial_\alpha f_0 -  \int \mathrm{dP} p^\alpha \partial_\alpha \left[\frac{\tau_R}{(u \cdot p)} p^\beta \partial_\beta f_0\right] + \int \mathrm{dP} p^\beta \partial_\beta \delta f^*_{(1)}, \\
    &= \mathcal{I_A} +\mathcal{I_B} + \mathcal{I_C}.
\end{aligned}    
\end{equation}
It is important to emphasize that the term $\mathcal{I_C}$ that depends on the $\delta f_{(1)}^*$ solely arises due to the momentum dependence of relaxation time and in general depends on $\delta\mu=\mu^*-\mu$, $\delta T=T^*-T$, and $\delta u_\mu=u_\mu^*-u_\mu$.
Using Eq.~(\ref{2.25}), we evaluate $\mathcal{A}$ as,
\begin{equation}
\begin{aligned}
    \mathcal{I_A} &= \int \mathrm{dP} p^\alpha \partial_\alpha f_0 \\
    &= - \int (u \cdot p) \Bigg[ \left(\frac{n}{\epsilon+P} - \frac{1}{(u \cdot p)}\right) p^m \nabla_m \alpha +\beta \frac{p^m p^n}{(u \cdot p)} \sigma_{mn} -\frac{\beta}{\epsilon+P} \left\{ p^m \nabla_n \pi^n_m - p^m \pi_{nm}\dot{u}^n\right\} \\
    & + \left\{ A_n - D_n (u \cdot p) \right\} \partial_\alpha n^\alpha + \left\{ A_\Pi - \left( D_\Pi + \frac{\beta}{(\epsilon+P)} \right) (u \cdot p)\right\} \pi^{\alpha \beta} \sigma_{\alpha \beta} \Bigg] f_0 \\
    &= -\frac{\beta I_{20}}{\epsilon+P} \pi^{\alpha \beta} \sigma_{\alpha \beta} + \left(D_n I_{20} - A_n I_{10}\right) \partial_\alpha n^\alpha + \left\{\left(D_\Pi + \frac{\beta}{\epsilon+P}\right) I_{20} - A_\Pi I_{10}\right\} \pi^{\alpha \beta} \sigma_{\alpha \beta}.
\end{aligned}    
\end{equation}
Using the definitions of $A_n$, $D_n$, $A_\Pi$ and $D_\Pi$ from Eq.~(\ref{m2.12}) and Eq.~(\ref{m2.13}) along with the definitions of the thermodynamic integrals introduced in Appendix \textbf{A}, we have the following relations:
\begin{align}
    \frac{\beta I_{20}}{\epsilon+P} &= \frac{3\beta}{4}, \\
    A_n I_{10} &= D_n I_{20} +1, \\
    A_\Pi I_{10} &= \left(D_\Pi + \frac{\beta}{\epsilon+P}\right) I_{20} - \frac{3\beta}{4}. 
\end{align}
With the above relations, we obtain $\mathcal{I_A} = -\partial_\alpha n^\alpha$. Similarly, evaluating $\mathcal{B}$ keeping terms up to second order in gradients,
\begin{equation}
\begin{aligned}
   \mathcal{I_B} &= -\int \mathrm{dP} p^\alpha \partial_\alpha \left[ 
 \frac{\tau_R}{(u \cdot p)} p^\beta \partial_\beta f_0 \right] = \int \mathrm{dP} p^\alpha \partial_\alpha \left\{ \tau_R \left( \frac{n}{\epsilon+P} - \frac{1}{(u \cdot p)} \right) p^\beta \nabla_\beta \alpha + \beta \frac{\tau_R p^m p^n}{(u \cdot p)}\sigma_{mn} \right\}f_0 \\
 &= \partial_\alpha \left[ \left( \frac{n}{\epsilon+P} K_{21} - K_{11} \right) \nabla^\alpha \alpha\right].
\end{aligned} 
\end{equation}
Using the expression of $\delta f_{(1)}^*$ as defined in Eq.~(\ref{2.15}) and the first order matching conditions Eq.~(\ref{2.18}) for a conformal system, we obtain $\mathcal{I_C}$ as,
\begin{equation}
    \mathcal{I_C} = \int \mathrm{dP} p^\alpha \partial_\alpha \delta f_{(1)}^* = -\partial_\alpha \left[ C\beta^2 I_{21} \nabla^\alpha \alpha \right].
\end{equation}
Now, we evaluate the zeroth moment of the ERTA collision term, and by using the definition of first-order number diffusion from Eq.~(\ref{re26}), we obtain,
\begin{equation}
\begin{aligned}
    \int \mathrm{dP} C[f] &= \mathcal{I_A} + \mathcal{I_B} + \mathcal{I_C} = -\partial_\alpha n^\alpha + \partial_\alpha \left[\left( \frac{n}{\epsilon+P} K_{21} - K_{11} \right) \nabla^\alpha \alpha - C\beta^2 I_{21} \nabla^\alpha \alpha\right], \\
    &= -\partial_\alpha n^\alpha +\partial_\alpha n^\alpha = 0.
\end{aligned}    
\end{equation}
Hence, we have verified that the zeroth moment of the ERTA collision kernel becomes zero where the extra part introduced due to the momentum dependence of the relaxation time is exactly canceled by the $\delta f_{(1)}^*$ term's contribution that we got from the first order matching condition. This ensures the number conservation law holds in this model. 
\subsubsection{Energy-momentum Conservation: $\partial_\mu T^{\mu\nu}=0$}
To verify the energy-momentum conservation in the current model, we need to show that the first momentum moment of the collision kernel is zero. To that end, we calculate the following
\begin{equation}
\begin{aligned}
    \int \mathrm{dP} p^\mu C[f] &= \int \mathrm{dP} p^\mu p^\alpha \partial_\alpha f_0 -  \int \mathrm{dP} p^\mu p^\alpha \partial_\alpha \left[\frac{\tau_R}{(u \cdot p)} p^\beta \partial_\beta f_0\right] + \int \mathrm{dP} p^\mu p^\beta \partial_\beta \delta f^*_{(1)}, \\
    &= \mathcal{I_D} +\mathcal{I_E} + \mathcal{I_F}.
\end{aligned}    
\end{equation}
Again, using Eq.~(\ref{2.25}), we obtain,
\begin{equation}
\begin{aligned}
    \mathcal{I_D} &= \int \mathrm{dP} p^\mu p^\alpha \partial_\alpha f_0 \\
    &= - \int (u \cdot p) \Bigg[ \left(\frac{n}{\epsilon+P} - \frac{1}{(u \cdot p)}\right) p^\mu p^m \nabla_m \alpha +\beta p^\mu \frac{p^\mu p^m p^n}{(u \cdot p)} \sigma_{mn} -\frac{\beta p^\mu}{\epsilon+P} \left\{ p^m \nabla_n \pi^n_m - p^m \pi_{nm}\dot{u}^n\right\} \\
    & + \left\{ A_n - D_n (u \cdot p) \right\}p^\mu \partial_\alpha n^\alpha + \left\{ A_\Pi - \left( D_\Pi + \frac{\beta}{(\epsilon+P)} \right) (u \cdot p)\right\} p^\mu \pi^{\alpha \beta} \sigma_{\alpha \beta} \Bigg] f_0 \\
    &= \left( -\frac{n I_{31}}{\epsilon+P} + I_{21}\right)\nabla^\mu \alpha +\left(\frac{\beta I_{31}}{\epsilon+P} -\frac{\beta I_{30}}{\epsilon+P}\right)u^\mu \pi^{\alpha \beta}\sigma_{\alpha \beta}+\frac{\beta I_{31}}{\epsilon+P} \left(\nabla_\alpha \pi^{\alpha \mu} - \pi^{\alpha \mu} \dot{u}_\alpha\right)\\
    &+ \left(D_n I_{30} - A_n I_{20}\right) \partial_\alpha n^\alpha + \left(\left(D_\Pi + \frac{\beta}{\epsilon+ P}\right)I_{30} - A_\Pi I_{20}\right) u^\mu \pi^{\alpha \beta} \sigma_{\alpha \beta}.
\end{aligned}
\end{equation}
Using Eq.~(\ref{m2.12}) and Eq.~(\ref{m2.13}) along with the definitions of the thermodynamic integrals, we find:
\begin{align}
    \frac{\beta I_{30}}{\epsilon+P} &= 3,\\
    \frac{\beta I_{31}}{\epsilon+P} &= -1, \\
    \frac{n I_{31}}{\epsilon+P} &= I_{21}, \\
    D_n I_{30} &= A_n I_{20}, \\
    D_\Pi I_{30} &= 1+ A_\Pi I_{20}
\end{align}
Using the above relations, we find that $\mathcal{I_D} = \pi^{\mu \alpha}\dot{u}_\alpha - \nabla_\alpha \pi^{\mu \alpha}$. By evaluating $\mathcal{I_E}$, we have 
\begin{equation}
\begin{aligned}
   \mathcal{I_E} &= -\int \mathrm{dP} p^\mu p^\alpha \partial_\alpha \left[ 
 \frac{\tau_R}{(u \cdot p)} p^\beta \partial_\beta f_0 \right],\nonumber\\
 &= \int \mathrm{dP} p^\mu p^\alpha \partial_\alpha \left\{ \tau_R \left( \frac{n}{\epsilon+P} - \frac{1}{(u \cdot p)} \right) p^\beta \nabla_\beta \alpha + \beta \frac{\tau_R p^m p^n}{(u \cdot p)}\sigma_{mn} \right\}f_0, \\
 &= \partial_\alpha \left[ \left( \frac{n K_{31}}{\epsilon+P} - K_{21}\right)(u^\mu \nabla^\alpha \alpha + u^\alpha \nabla^\mu \alpha) \right]+ \partial_\alpha\left[2\beta K^{(1)}_{42} \sigma^{\mu \alpha}\right]. 
\end{aligned} 
\end{equation}
By employing Eq.~(\ref{re25}), the last term  in the above equation can be written as $\partial_\alpha \pi^{\mu \alpha} = -\dot{u}_\alpha \pi^{\mu \alpha} + \nabla_\alpha \pi^{\mu \alpha}$,
\begin{equation}
    \mathcal{I_E} = \partial_\alpha \left[ \left( \frac{n K_{31}}{\epsilon+P} - K_{21}\right)(u^\mu \nabla^\alpha \alpha + u^\alpha \nabla^\mu \alpha) \right]- \dot{u}_\alpha \pi^{\mu \alpha} + \nabla_\alpha \pi^{\mu \alpha}.
\end{equation}
Similar to the previous subsection, using the expression for $\delta f_{(1)}^*$, which solely arises due to the momentum dependence of thermal relaxation time, obtained from the matching condition, we obtain
\begin{equation}
    \mathcal{I_F} = \int \mathrm{dP} p^\mu p^\alpha \partial_\alpha \delta f_{(1)}^* = -\partial_\alpha \left[ C\beta^2 I_{31} (u^\mu \nabla^\alpha + u^\alpha \nabla^\mu \alpha) \right].
\end{equation}
Finally, we estimate the first moment of the collision kernel as,
\begin{equation}
    \begin{aligned}
        \int \mathrm{dP} p^\mu C[f] = \mathcal{I_D} +\mathcal{I_E} + \mathcal{I_F} = \partial_\alpha \left[ \left\{ \left( \frac{n K_{31}}{\epsilon+P} - K_{21}\right) -C\beta^2 I_{31}\right\} (u^\mu \nabla^\alpha + u^\alpha \nabla^\mu \alpha) \right].
    \end{aligned}
\end{equation}
Using the fact that, 
\begin{equation}
    \left( \frac{n K_{31}}{\epsilon+P} - K_{21}\right) =C\beta^2 I_{31}, 
\end{equation}
we obtain,
\begin{equation}
    \int \mathrm{dP} p^\mu C[f] = 0.
\end{equation}
Hence, we have verified that the first moment of the ERTA collision term vanishes, thereby guaranteeing the conservation of energy-momentum tensor in the current model. We find that the contribution to the distribution function from the first order matching condition exactly cancels out the remaining non-zero part due to the momentum-dependent relaxation time. In general, we note that the form of $\mu^*=\mu+\delta \mu$, $T^*=T+\delta T$, and $u_\mu^*=u_\mu+\delta u_\mu$ are important in obtaining the conservation equations. The definitions of $\delta \mu$, $\delta T$, and $\delta u_\mu$ incorporate the counter terms that arise due to the momentum dependence of relaxation time for the satisfaction of the conservation equations. The definition of these quantities in the first order in gradients are defined in Eq.~(\ref{2.18}). For the present system, we have further defined $\delta u^\mu$ in terms of thermodynamic integrals (using distribution function) as,
\begin{equation*}
\begin{aligned}
u_\mu^*=u_\mu+ \frac{1}{\beta I_{31}^+}\left( \frac{n}{\epsilon+P}K_{31}^+ - K_{21}^- \right) \nabla_\mu \alpha. 
\end{aligned} 
\end{equation*}
However, one needs to define these quantities until the second order for the third-order hydrodynamic estimations. For a simplified system of massless particles at vanishing chemical potential, these can be defined as follows~\cite{Dash:2023ppc},
\begin{equation*}
\begin{aligned}
   &u_\mu^* = u_\mu +  \frac{5 K_{32}}{T(\epsilon+P)^2}\bigg(\pi^{\mu\nu}\dot{u}_\nu-\nabla_\nu\pi^{\mu\nu}-\pi^{\alpha\beta}\sigma_{\alpha\beta}u^\mu\bigg)+ \frac{2 L_{32}}{T(\epsilon+P)}\bigg(2\sigma^{\mu\nu}\dot{u}_\nu+\nabla_\nu\sigma^{\mu\nu}+\sigma^{\alpha\beta}\sigma_{\alpha\beta}u^\mu\bigg),\nonumber\\
   &T^* = T +  \frac{5 K_{32}}{3(\epsilon+P)^2}\pi^{\mu\nu}\sigma_{\mu\nu}+\frac{1}{\epsilon+P}\Big(L_{32}-\frac{L_{42}}{3T}\Big)\sigma^{\mu\nu}\sigma_{\mu\nu}.
\end{aligned} 
\end{equation*}
The contribution from these quantities becomes essential 
to ensure the conservation laws within the framework of third-order hydrodynamics, which is beyond the scope of the present analysis. }



 \bibliographystyle{apsrev4-1}
\bibliography{biblio.bib}

\end{document}